\documentclass[10pt,sigconf,letterpaper,authorversion,nonacm]{acmart}

\usepackage[english]{babel}
\usepackage{blindtext}
\usepackage{color}
\usepackage{xurl}
\usepackage{amsmath,amsfonts}
\usepackage{algorithmic}
\usepackage{textcomp}
\usepackage{svg}
\usepackage{xcolor}
\usepackage{graphicx}
\usepackage{graphics}
\usepackage{float}
\usepackage{cclicenses}
\usepackage{xspace}
\usepackage{subfigure}
\usepackage{caption}
\usepackage{makecell}
\usepackage{multirow} 
\usepackage{mathtools}
\usepackage[normalem]{ulem}
\usepackage{array}
\usepackage{svg}
\usepackage{booktabs}
\usepackage{multirow}
\usepackage{pgfplots}
\usepackage{tikz}
\usepackage{afterpage}
\usepackage{tabulary}
\usepackage{longtable}
\usepackage{float}
\usetikzlibrary{intersections}
\usetikzlibrary[shapes,arrows]
\usetikzlibrary{shapes.geometric}
\usepgfplotslibrary{fillbetween}
\usetikzlibrary{positioning}
\RequirePackage{filecontents}

\acmYear{2018}
\copyrightyear{2018}
\setcopyright{acmcopyright}
\acmConference{CoNEXT '18}{December 4--7, 2018}{Heraklion/Crete, Greece}
\acmPrice{TBA}
\acmDOI{TBA}
\acmISBN{TBA}

\makeatletter
\def\@copyrightspace{\relax}
\makeatother

\setcopyright{none}
\renewcommand\footnotetextcopyrightpermission[1]{} % removes footnote with conference information in first column
\newif\ifcomments
\commentstrue
\ifcomments

\commentsfalse

\ifcomments
\newcommand{\mas}[1]{{\color{red}#1}}
\else
\newcommand{\mas}[1]{{#1}}
\fi

\newcommand{\msd}[1]{{#1}}
 
\newcommand{\sx}[1]{(\S\ref{#1})}

\begin{document}
\title{A Tale of Frozen Clouds: Quantifying the Impact of Algorithmic Complexity Vulnerabilities in Popular Web Servers}

% \subtitle{Paper \# XXX, XXX pages}
\author{Masudul Hasan Masud Bhuiyan}
% \authornote{Note}
\orcid{0000-0002-7090-4334}
\affiliation{%
  \institution{CISPA Helmholtz Center for Information Security}
}
\email{masudul.bhuiyan@cispa.de}

\author{Cristian-Alexandru Staicu}
% \authornote{Note}
% \orcid{0000-0002-7090-4334}
\affiliation{%
  \institution{CISPA Helmholtz Center for Information Security}
}
\email{staicu@cispa.de}

\begin{abstract}
Algorithmic complexity vulnerabilities are a class of security problems that enables attackers to trigger the worst-case complexity of certain algorithms. Such vulnerabilities can be leveraged to deploy low-volume, asymmetric, {CPU-based} denial-of-service (DoS) attacks. Previous work speculates that these vulnerabilities are more dangerous in certain web servers, like Node.js, than in traditional ones, like Apache. We believe it is of utmost importance to understand if this is indeed the case or if there are ways to compensate against such problems using various deployment strategies. To this end, we study the resilience of popular web servers against CPU-based DoS attacks in four major cloud platforms under realistic deployment conditions. We find that there are indeed significant differences in how various web servers react to an attack.
However, our results suggest a more nuanced landscape than previously believed: while event-based systems tend to recover faster from DoS in certain scenarios, they also suffer the worst performance degradation overall. Nevertheless, in some setups, Apache performs worse than event-based systems, and there are cloud platforms in which all the considered servers are seriously exposed to the attack. We also find that developers can harden their servers against CPU-based DoS attacks by increasing the number of server instances running in parallel. This, in turn, can lead to an increased cost of operation or a slight degradation of performance in non-DoS conditions.
\end{abstract}

\maketitle
% \section{Introduction}
\section{Introduction}
In a classical, distributed denial-of-service (DDoS) attack, the adversary dedicates significant resources like thousands or millions of machines, possibly as part of a botnet, to attack the target system. Recently, Cloudflare mitigated one of the largest DDoS attacks with 17.2 million requests per second that originated from over 20,000 bots\footnote{\url{https://blog.cloudflare.com/cloudflare-thwarts-17-2m-rps-ddos-attack-the-largest-ever-reported/}}. In contrast, CPU-based DoS attacks can be carried away by a single adversary with limited bandwidth and computing resources, achieving a similar impact. 
%Because of this stealth mode, ACV attacks are more lethal to the target system. % Than what?
This type of attack is enabled by an easy-to-trigger, slow computation, available on the server-side, \textit{which the attacker can trigger repeatedly}. While such CPU-heavy computations can be introduced by careless developers, e.g., writing a file to disk using synchronous API in event-based systems, in recent years, we have seen an increased interest in studying algorithmic complexity vulnerabilities as a building block for CPU-based DoS attacks~\cite{staicu_freezing_2018, DavisCSL18, davis2021using, bai_runtime_2021, davis2019rethinking, turonovacounting, barlas2022exploiting}.
These types of vulnerabilities enable an adversary to trigger the worst-case performance of a target algorithm by providing carefully crafted inputs.

One example of algorithmic complexity vulnerability is regular expression denial of service (ReDoS), where the attacker exploits a backtracking-based algorithm in the regular expression engine by crafting adversarial inputs. For example, it takes 15 seconds for Node.js engines to match $/(a+)+b/$ against a sequence of 30 $a$ characters~\cite{staicu_freezing_2018}. Practitioners rely on simple rules of thumb to judge ReDoS: if a regular expression can be used to cause a one-second slowdown, using reasonable-sized input, it is considered vulnerable\footnote{\href{https://github.com/salesforce/tough-cookie/issues/92\#issuecomment-328173164}{Node Security Project considers every regular expression that can be used to trigger a slowdown larger than one second as vulnerable.}}.
There is an extensive body of recent work studying this security problem~\cite{ojamaa2012assessing, case_davis_2017, staicu_freezing_2018}, showing that it affects web applications in production~\cite{staicu_freezing_2018}, and proposing tools for detecting and remedying it~\cite{davis_using_2021,static_wustholz_2017,bai_runtime_2021,liu_revealer_2021,shen_rescue_2018}.

However, we believe that the relation between algorithmic complexity vulnerabilities and CPU-based DoS attacks is not fully understood by the community, and that there are a lot of speculations, unsupported by solid empirical evidence. First, both researchers, e.g., Davis et al.~\cite{case_davis_2017} and Staicu and Pradel~\cite{staicu_freezing_2018}, and practitioners\footnote{\url{https://snyk.io/blog/redos-and-catastrophic-backtracking/}}\footnote{\url{https://nodejs.org/en/docs/guides/dont-block-the-event-loop/}} claim that algorithmic complexity vulnerabilities, and ReDoS in particular, are more severe in asynchronous, event-based systems than in traditional web servers. Supposedly, this is due to the single-threaded nature of these runtimes. Second, practitioners\footnotemark[2], but also related academic work~\cite{DavisWL18,staicu_freezing_2018}, mostly consider payloads larger than one second as problematic. More generally, Davis et al.~\cite{DavisWL18} postulate that users can set a clear timeout threshold for slow operations, below which requests are considered harmless.
We challenge these assumptions by studying the resilience of modern web servers to CPU-based DoS attacks, under realistic deployment conditions. 
%We do not limit our study to event-based versus one-thread-per-client architectures, but also include systems relying on lightweight threads, i.e., goroutines in Go. 

%While previous work studies algorithmic complexity vulnerabilities in individual systems or programming languages: Node.js~\cite{case_davis_2017, ojamaa2012assessing}, Java~\cite{static_wustholz_2017} or Apache/PHP~\cite{crosby_denial_2003,meng_rampart_2018}, 
To the best of our knowledge, we are the first to perform a comparative study of the servers' landscape to assess the impact of algorithmic complexity vulnerabilities in widely-used web servers.  We believe that our measurements can help the community better understand these types of vulnerabilities, and at the same time, shed light on an unexplored tradeoff involved in choosing the design of web servers.
%, it does  not discuss in detail how an attacker may use such vulnerabilities to mount a CPU-based DoS attack. Crosby et al.~\cite{crosby_denial_2003} and Meng et al.~\cite{meng_rampart_2018} study to some extent the relation between algorithmic complexity vulnerabilities and denial-of-service attacks, but they do so for traditional web server architectures, e.g., PHP. 
%To the best of our knowledge, our work is the first to focus on the relation between the architectural design of a system and its vulnerability to CPU-based DoS attacks. 
Specifically, we answer the following research questions: 
    \begin{itemize}
    %   \item \textbf{Q1:} How do different web servers \underline{deployed locally} respond to CPU-based DoS attacks? Are some servers more exposed to this attack than others? \sx{res:local}
      \item \textbf{Q1:} How do different web servers {deployed in the cloud} perform under CPU-based DoS attacks? Are some servers more exposed to this attack than others? \sx{res:cloud}
      \item \textbf{Q2:} Is it possible to mitigate CPU-based DoS attacks using different deployment strategies? \sx{res:mitigation}
      
      \item \textbf{Q3:} Can slowdowns smaller than one second be leveraged for low-bandwidth DoS attacks against web servers deployed in the cloud? \sx{res:threshold}
    \end{itemize}

Answering these questions brings up a crucial methodological challenge: how can we uniformly set up different servers to observe them under the same attack so that no particular server faces unjustified advantages or disadvantages over the others. This challenge stems from the availability of diverse deployment strategies such as different hardware, load balancers, intrusion detection systems, etc. 
%One can also choose between different virtualization options, operating systems, hardware components, and so on. And 
All these different choices may have a significant effect on the performance of the server, influencing our measurement study. 

% We address this challenge using a two-fold strategy: (i) for our local setup, we choose the out-of-the-box setup for each web server, (ii) for our cloud setup, we choose different platform-as-a-service (PaaS) solutions to deploy our study on. 

We address this challenge by using platform-as-a-service (PaaS) solutions to deploy testbeds for our study. PaaS platforms provide a ready-to-use software and hardware stack with minimal configuration options, i.e., developers' only tasks are to provide the server's source code and choose a pricing model. 
In this way, we use a fair testbed, ensuring that we do not inadvertently advantage certain systems over others, and at the same time, study straight-forward deployment strategies available to developers.

% Underlying our measurements study, there is a novel methodology for quantifying the impact of denial-of-service attacks. We propose remotely benchmarking different web servers under simulated attacks, and collecting newly-introduced metrics, such as attacker's gain and throttled time. We apply this methodology to both the considered setups and compare the response of different systems to CPU-based DoS attacks.

Underlying our study, there is a novel methodology for quantifying the impact of denial-of-service attacks. We benchmark different web servers under simulated attacks using the most popular benchmarking tools used by the practitioners and proposed two new metrics, such as \emph{Attacker's gain} and \emph{Throttled time} to summarise the effects of CPU-based DoS attacks under different attack scenarios.
We apply this methodology to both the considered setups and compare the response of different systems to CPU-based DoS attacks. It is worth noting that we use our newly proposed metrics to summarize and combine multiple attack scenarios for quantitative comparison. Both metrics are calculated from the throughput, and together with latency, they portray the effects of CPU-based DoS attacks on different web servers. 

% local
% Our local experiments confirm that in their out-of-the-box setup, event-based web servers like Node.js are significantly more vulnerable to CPU-based DoS attacks than the other considered systems. A low-bandwidth attack can have a dramatic effect both on the throughput of the server and on the user's perceived latency. For example, a five seconds attack with 100 requests per second bandwidth, and 500 milliseconds payloads yields a Node.js server completely unresponsive for 25 seconds and increases the latency of 1,650  concurrent requests above 10 seconds. In contrast, Apache web servers show no measurable decrease in throughput under the same attack.
% cloud
% Our cloud experiments reveal a more complex picture: the difference between servers is significantly reduced when they are deployed under realistic conditions, i.e., on PaaS platforms. Moreover, we observe that in several cloud setups the performance of event-based systems is comparable, or even better, than Apache's performance. 
%Nevertheless, there are still important differences among servers,  especially when more powerful, paid instances are used.

Our results reveal a more complex picture than previously believed. Even though event-based web servers like Node.js are significantly more vulnerable to CPU-based DoS attacks than the other considered systems in most cases, the difference between servers is significantly reduced when they are deployed under constrained conditions, i.e., on the free tier. Moreover, we observe that in several cloud setups, the performance of event-based systems is comparable, or even better, than Apache's performance. 

% mitigation
Additionally, we show that developers have multiple configuration options to increase the resilience of their single-threaded runtimes to CPU-based DoS. They can increase the number of preforked workers, an option available in both the considered event-based servers. Doing so significantly reduces the susceptibility of these systems to CPU-based DoS attacks, but it may affect the server performance in non-DoS conditions. They can also deploy multiple powerful machines, but that will also incur additional costs.
%, and can lead to a degradation of performance for the non-attack conditions. 
Concretely, we show that by carefully configuring this parameter, one can achieve an attack response for Node.js comparable to that of multi-threaded systems like Tomcat.

% threshold
Finally, our measurements are a warning to the community to also consider sub-second slowdowns as potential vulnerabilities. We show that a 50 milliseconds slowdown can be weaponized against well-configured systems in the considered cloud setups.

In summary, this work makes the following contributions:
\begin{itemize}

\item We provide empirical evidence about the fundamental differences between the responses of different web servers to CPU-based DoS attacks. We find that event-based servers tend to exhibit a very sudden drop in quality of service, while one-thread-per-client ones suffer more gradual performance degradation.

\item We discuss developers' options for configuring their cloud applications to be more resilient against CPU-based DoS attacks. We propose further ways to reduce the differences between server architectures.

\item We show that the threshold used by practitioners for judging algorithmic complexity vulnerabilities is too high, and we advocate for case-by-case consideration.

% \item We propose a novel methodology for quantifying the damage incurred by CPU-based DoS attacks. We define metrics like attacker's gain and throttled time to summarize the server's performance degradation over time.

\end{itemize}

\iffalse
\begin{itemize}
  \item 
  \item Does the architectural design of a server make it more prune to ACV attacks?
  \item Is it possible to mitigate ACV attacks using different deployment strategies for the same server architecture?
  \item How do different architectures perform under ACV attacks in different cloud platforms?
  \item Can a state-of-the-art DDoS defense system mitigate ACV attacks in different server architectures? 
\end{itemize}
\fi
\section{Background}

In this section, we introduce typical web server architectures used by practitioners, and the platform as a service paradigm. 

\subsection{Web server architectures}

During the years, many architectures were proposed for building web servers, each promising the best performance. It is beyond the scope of this paper to survey all of them, and thus, we limit our discussion to the most important classes, highlighting popular servers implementing them.

\paragraph{\textbf{One thread per client}}
% E.g., Spring, PHP with Apache
In this architecture, each incoming request gets mapped to a single thread. The threads are sometimes managed by a thread pool, to avoid the high cost of spawning new threads on demand. The main advantage of this architecture is its simplicity, while the main disadvantage is poor performance in the presence of long-lasting requests. This architecture is the most widely adopted one, with prominent examples including Apache\footnote{\url{https://httpd.apache.org/docs/2.2/mod/worker.html}} and Tomcat\footnote{\url{https://tomcat.apache.org/tomcat-7.0-doc/jdbc-pool.html}}.
% \vspace{-5mm}
\paragraph{\textbf{One process per client}}
In this architecture, each request is mapped to a single process. A pool of processes is often spawned when the server starts and is sometimes referred to as \textit{preforked} workers or processes. The advantages of this architecture are a more clear separation of user requests than in the previous architecture and the avoidance of thread-safety bugs.
The most popular server implementing this strategy is Apache in \texttt{prefork} mode\footnote{\url{https://httpd.apache.org/docs/2.4/mod/prefork.html}}.
% E.g., ?
% \vspace{-5mm}
\paragraph{\textbf{Single-threaded, async I/O}}
% E.g., Node.js
Managing the life cycle of processes or threads, and context switching between them, as proposed by previous architectures, comes at a high cost. Hence, in recent years, we have seen an increased interest in single-threaded, event-based architectures. In this case, the server handles all the requests in a single thread and offloads slow computation, e.g., input/output operations, to a thread pool. The most widely adopted server implementing this mode of operation is Node.js.
% \vspace{-5mm}
\paragraph{\textbf{Preforked workers, async I/O}}
% E.g.: Gunicorn with Django, Node.js to some extent
Practitioners also use a hybrid between two of the architectures above: multiple single-thread, event-based workers are preforked on the same machine, and incoming requests are distributed to the worker with the lowest load at each moment. Gunicorn\footnote{\url{https://github.com/benoitc/gunicorn}} for Python uses this default mode of operation, and recently, Node.js introduced the \texttt{Cluster API} to enable it as well. In this paper, we sometimes use the term \textit{event-driven architectures} to refer to the last two server architectures described above.
% \vspace{-3mm}

\paragraph{\textbf{Light threads}}
A different approach for reducing the overhead caused by multiple threads is multiplexing several requests on the same thread. This can be done using so-called light threads, e.g., goroutines. A prominent server example of this architecture is the built-in Go server~\cite{benoitc_github}.

As seen from the examples above, modern web servers sometimes offer configuration options to choose between different architectures. For example, Apache offers most of the above modes of operation\footnote{\url{https://en.wikipedia.org/wiki/Apache_HTTP_Server\#Performance}}.

% \paragraph{\textbf{Light threads}}
% % E.g., Go
% A different approach for reducing the overhead caused by multiple threads is to multiplex several requests on the same thread. This can be done using so-called light threads, e.g., goroutines. A prominent server example for this architecture is the built-in Go server\footnote{\url{https://golang.org/doc/articles/wiki/}}.

% As seen from the examples above, modern web servers sometimes offer configuration options to choose between different architectures. For example, Apache offers most of the above modes of operation\footnote{\url{https://en.wikipedia.org/wiki/Apache_HTTP_Server\#Performance}}.
% Say that it is not black and white and even traditional servers are adopting new modes of operation: https://en.wikipedia.org/wiki/Apache_HTTP_Server#Performance
% \vspace{-3mm}
\subsection{Platform as a service}
% only discuss things that are relevant to us
% promises: ready-to-use infrastructure, you do not need to rent an entire server, concentrate on code
Platform as a service (PaaS) is a cloud computing paradigm in which developers seemingly deploy web services as software bundles without worrying about the underlying infrastructure. The PaaS platform handles the setup of the actual machine and its software stack, and provides ready-to-use \emph{instances} for deploying the bundles on. 

% explain how it works: deploy by committing on a repo
\iffalse
The publishing model of PaaS platforms is very straightforward: the developer selects the type and number of instances, provides the source code using code repository or code bundles. It is also possible to use containers for deploying the server. From there, the platform initializes the necessary instance(s) with the code, enabling instantaneous deployments. The code must often use one of the programming languages or frameworks supported by the platform.

% explain pricing model
The pricing model usually depends on the computation capabilities of the deployed instances and their actual number. Developers can decide to rent a fixed number of instances for a given project, in accordance with their expected traffic, or they can opt for autoscaling, in which case, new instances are added on-demand to match any increase in user traffic. 
Often, PaaS platforms also provide free of cost, resource-constrained instances, to enable developers to experiment with the platform at zero cost. However, these instances are not recommended for production as they exhibit poor performance.
\fi

% benefits
PaaS paradigm promises several benefits: it allows developers to concentrate on writing code instead of spending time managing the infrastructure, and it reduces costs when compared to the more traditional modes of operation. Instead of providing an entire physical machine for each user, in the PaaS case, a single machine can host multiple virtual ones, each running a PaaS instance allocated to a different user. 

% cons: less control; you need to accept most of the configs provided by the platform
The main drawback of the PaaS paradigm is the loss of control for developers. Most of the configurations, such as which operating system version or load balancer to use, are not under the developer's control but decided by the platform. This perceived drawback makes PaaS systems an attractive target for our measurements study: they enable us to easily deploy the web servers we want to study in the cloud and observe their behavior under realistic deployment conditions.

\section{Methodology}
\label{sec:methodology}

As discussed in introduction, the two core challenges for our measurements study are: how to simulate equivalent attacks against different web servers and how to measure the attack's impact. 
%In Figure~\ref{fig:overview}, we show a high-level representation of our methodology that tackles these two hard problems, and 
Below, we discuss in detail our novel methodology that tackles these two challenges. We first define the attacker's capabilities and how a DoS attack affects the users of a system \sx{subsec:tm}. We then show how to specify reusable CPU-based DoS attacks that can be simulated against multiple systems \sx{subsec:attspec}. Finally, we discuss a set of metrics to be collected in the empirical study \sx{subsec:metrics}.

\subsection{Threat model}
\label{subsec:tm}
We assume that two users are interacting with a system simultaneously: the attacker and the victim. The attacker's objective is to reduce the server's quality of service by sending several well-crafted requests to the server. In the ideal case, the victim should not perceive any degradation of the server's quality of service. Using this intuition, we can define a system's resilience against denial-of-service attacks:

\begin{definition}
Let an observer perform two sets of measurements: ${M}_1$ under attack conditions, and ${M}_2$ under non-attack conditions. We say that a system is \textbf{DoS-noninterferent}, if given the pair $\{{M}_1, {M}_2\}$, the observer cannot reliably choose the measurements done under attack conditions.
\end{definition}

 DoS-noninterference is a noble goal for a system, but we notice that if the attack is powerful enough, most systems do not achieve it in practice. Hence, an observer can obtain information about the server's performance by measuring certain properties of the served requests, e.g., response time.
We propose quantifying the server's performance degradation by repeatedly observing the difference between the measurements under attack ($M_1$) and non-attack ($M_2$) conditions.

\subsection{Attack specification}
\label{subsec:attspec}

We remind the reader that our objective is to simulate equivalent CPU-based DoS attacks against multiple systems. Thus, we need to carefully control a set of attack parameters.

\paragraph{\textbf{Payload size}}
Each request made to a web server consumes a certain CPU time budget on the server. Hence, for an individual request, we define the \emph{payload size} as the CPU time consumed on the server-side for processing that request, and we measure it in milliseconds. Measuring this parameter is not trivial, and it is highly dependent on the exact analyzed system. In our experimental setup, we describe how we precompute this value for the considered systems.

\paragraph{\textbf{Bandwidth-bound attacker}} We assume an attacker cannot send arbitrary many requests to a server. This is because production-ready applications usually deploy some form of rate-limiting or intrusion detection systems that constrain the attacker's capabilities. Hence, we limit the number of requests sent by the attacker to an upper bound value called \emph{attack's bandwidth}, and we measure it in requests per second. 
%The attacker should also have predictable behavior, e.g., during two independent attacks, it should use the \msd {same number of concurrent connections.}

% same number of simultaneously open connections.
% measured in req/second

% measured in milliseconds

\paragraph{\textbf{Attack window}} We assume the attacker sends malicious requests in a limited time interval called \emph{attack window}, which we specify in seconds.  Shan et al.~\cite{shan_tail_2017} show that denial-of-service attacks consisting of short bursts, followed by longer cool-down periods are effective against web applications. In this work, we aim to study the system's behavior both during the attack and in the recovery phase.
% measured in seconds

Using the above parameters, we define a CPU-based denial-of-service attack as the tuple $\mathcal{A} = (p, b, w)$, where $p$ is the payload size, $b$ is the bandwidth, and $w$ is the attack window. We argue that by carefully controlling these parameters, one can instantiate equivalent attacks against different systems.

\subsection{Measurements}
\label{subsec:metrics}

% Local vs. remote measurements: pros and cons
As discussed in Section~\ref{subsec:tm}, we propose measuring the server's performance remotely. More precisely, the observer machine performs several requests and records relevant parameters for the served responses. While this way of measuring may appear noisy at first, we argue that it reflects the real performance degradation of a system, as perceived by the end-user. We believe that realistic measurements should include all the factors that can influence the user experience, beyond the actual throughput measured on the server's end, e.g., load balancers, firewalls, network caches. This is also in line with the mode of operation of popular benchmarking tools used by developers. In fact, we build our measurements infrastructure on top of wrk2\footnote{\url{https://github.com/giltene/wrk2}}, a widely-used open-source benchmarking tool for web applications. 
We describe below the most important measurements we collect and a couple of relevant parameters for our study.

\paragraph{\textbf{Measurement rate}} We call the number of measurement requests sent by an observer in a given time interval, \emph{measurement rate}, and we measure it in requests per second. 
To avoid negatively influencing the measurements, the observer must strive to send measurement requests with minimal CPU-load and avoid caching, e.g., by appending a random value to the request. They must also ensure that the sum of the measurement rate and the attack's bandwidth is below the throughput of the server in non-attack conditions. Otherwise, instead of performing a measurements study, one would exhaust the bandwidth of the server.

\paragraph{\textbf{Measurement window}} The time interval in which the observer sends measurement requests is called \emph{measurement window} $\mathcal{W}$, and we measure it in seconds. The measurement window should ideally include the attack window but not perfectly overlap. That is because we want to observe the server's behavior in non-attack conditions, during the attack, and in the recovery phase. 

\paragraph{\textbf{Throughput}} We call \emph{throughput} $\mathcal{T}(t)$ the number of requests an observer receives back from the server in a given second. We note that in the DoS-noninterference conditions, and when the processing time for a request is constant, the throughput should be equal to the measurement rate, which we sometimes call \emph{expected throughput} $\mathcal{E}(t)$. If these conditions are not met, the expected throughput should be calculated by repeatedly observing the server's throughput in non-attack conditions. We point the reader's attention to the fact that both $\mathcal{T}(t)$ and $\mathcal{E}(t)$ are discrete functions whose domains are limited to seconds in the measurement window.

\paragraph{\textbf{Latency}} For a given request, we call \emph{latency} the time interval between the moment it was sent from the observer's machine until its response was received back. We note that both latency and throughput have a slightly different meaning than previously defined in the literature: instead of measuring these two values on the server-side, we measure them at the observer's end. Hence, they do not directly measure the performance of the server on all the served requests, but only on the ones served to the observer. We argue that this is a more adequate way to measure a server's performance, in line with how practitioners benchmark their servers.

\begin{figure}
    \centering
    \includegraphics[width=\linewidth, keepaspectratio=true]{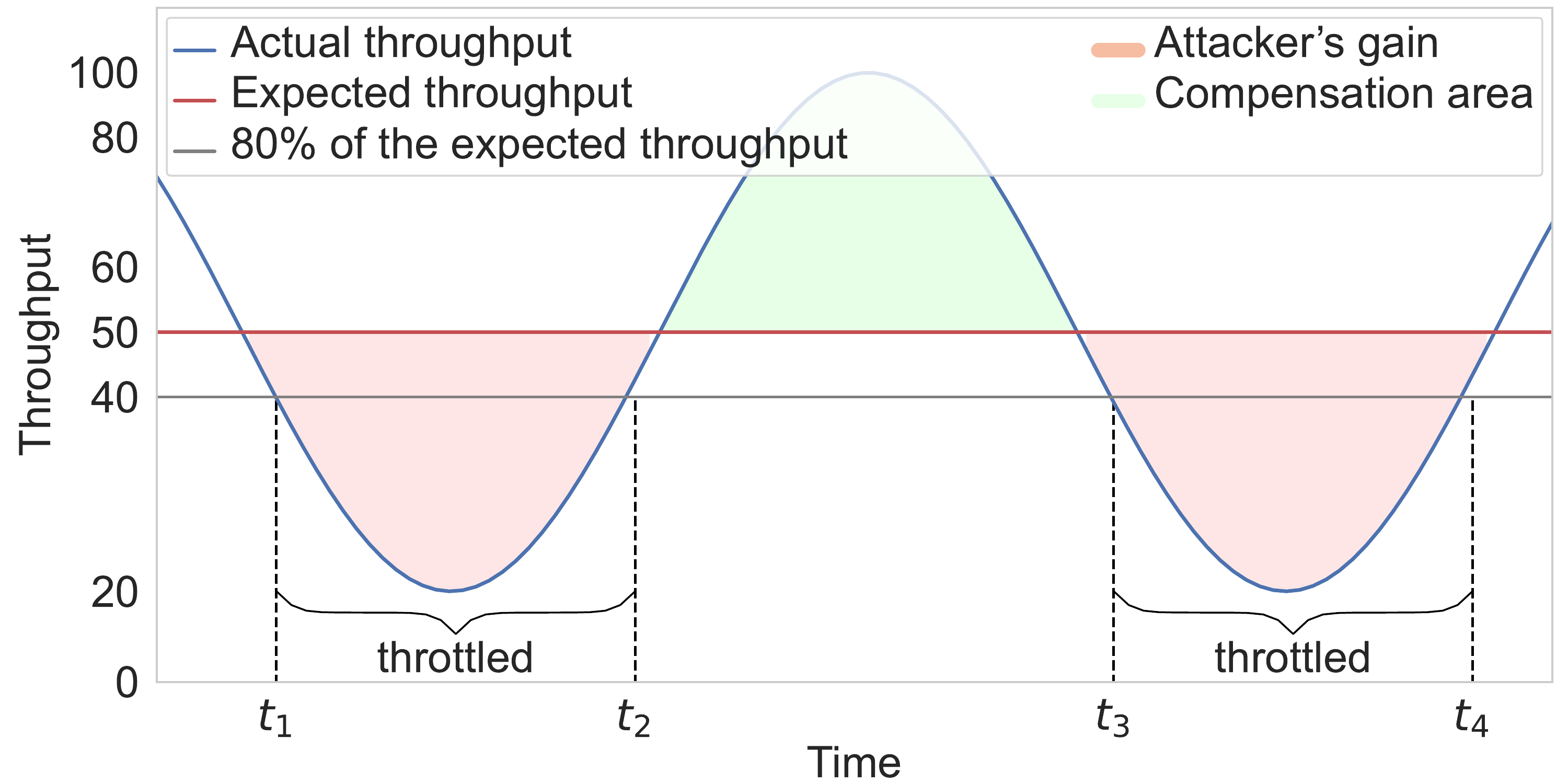}
    \caption{Visual representation of the throughput gain}
    \label{fig:t_gain}
\end{figure}

\paragraph{\textbf{Attacker's gain}}
\label{attacker's_gain}
% explain why we need this metrics
The two previously discussed metrics are enough for analyzing the behavior of a single server under a single attack. \msd{We observe that multiple benchmarking tools used by developers output these two values, often in the form of time series.} However, since in our study we aim to observe the effect of certain parameters on the attack's performance, we need to define some metrics that can capture the server's performance degradation in a single scalar value. \msd{Without these, one has to look into thousands time series graphs to assess the impact of an attack in different settings over multiple runs. Thus, for a large-scale comparison study that we aim to present in this work, defining new metrics is essential to summarise the degradation of the server's quality of service under different attack scenarios.}
Let us consider the actual server's throughput (blue line) in Figure~\ref{fig:t_gain}, and the expected throughput (red line) as 50 requests per second. As can be observed, the throughput seriously degrades reaching less than 20 requests per second for some time, followed by a compensation phase in which more requests are served than the expected throughput, and finally going through a second degradation phase in which the throughput goes below 40 requests per second. We assume that the depicted throughput graph is caused by two adjacent attacks. We propose calling the red area in the figure \emph{Attacker's gain}. This area shows the deviation of the actual throughput from its expected value, and it captures the perceived degradation in the quality of service, as measured by the observer. For example, at moment $t_1$, the observer expects to receive 50 requests back from the server, but only 40 are actually received. 
%Hence, we conclude that the server's quality of service decreased. 
Below, we introduce a rigorous definition for this intuition. 

Given the expected throughput $\mathcal{E}(t)$ of a server, and its actual throughput under attack $\mathcal{T}(t)$, we compute two polynomials $f(t)$ and $g(t)$ that interpolate the values in $\mathcal{E}(t)$ and $\mathcal{T}(t)$, respectively. We define the attacker's gain as:

\begin{equation}
    G = \int f(t) - \int \overline{g}(t)
\end{equation}

Where  $\overline{g}(t)$ is the actual throughput $g(t)$ upper bounded by the expected throughput $f(t)$:

\begin{equation}
    \overline{g}(t) = \begin{cases}
  g(t) & \text{$\forall \; t, s.t. \; g(t) < f(t)$}\\
      f(t) & \text{otherwise}
    \end{cases} 
\end{equation}

Without bounding $g(t)$, the difference between the two quantities will often be zero, because servers tend to recover after the attack stops. More precisely, if there are no requests dropped by the server, then $\int f(t) - \int {g}(t) = 0$. That is because some requests are served after a long time, but served nevertheless, i.e., the ones in compensation area in Figure~\ref{fig:t_gain}.
%However, $\overline{g}(t)$ captures the decrease in quality of service perceived by the user.
% explain it on the graph

\paragraph{\textbf{Throttled time}}
\label{throttled_time}
% explain why we need this metric
Attacker's gain captures the attack's effect on the throughput over time, but it does not capture how large the attack's impact is in its worst moments. That is, two areas of the same size can have very different shapes: one that is very high in amplitude and one that is rather stretched over a long period. Thus, we propose a metric that quantifies the amount of time the server's throughput was below a given threshold. By computing this metric for different thresholds, we can see both for how long the attack's effect lingered and how bad the peak of the attack was. In general, we consider the threshold as a percentage of the expected throughput.

% define it
Given a throughput $\mathcal{T}(t)$ and a percentage $p$, we define throttled time as:

\begin{equation}
   T_p = |\{x \in \mathcal{W} \; | \; \mathcal{T}(x) < p*\mathcal{E}(x)\}|
\end{equation}

% explain it on the graph
We say that "the throttled time at $p$ is $T_p$". For example, in Figure~\ref{fig:t_gain}, the throttled time at 80\% is $T_p = (t_2-t_1)+(t_4-t_3)$, i.e., in those time intervals the throughput is below 80\% of the expected throughput.

\iffalse
\begin{figure}
    \centering
    \includegraphics[width=4cm]{mock-figs/gain.png}
    \caption{Visual representation of the throughput gain}
    \label{fig:t_gain}
\end{figure}

\paragraph{Throttled time}

\begin{figure}
    \centering
    \includegraphics[width=4cm]{mock-figs/tt.png}
    \caption{Throttled time}
    \label{fig:t_time}
\end{figure}
\fi

We warn the reader about the dangers of using a single metric to quantify something as complex as servers' performance degradation. Looking at any of the metrics above, in isolation, is not sufficient.  For example, analyzing \emph{Attacker's gain}, without complementing it with throttled time or latency, may lead us to false conclusions. Hence, in the results section, we discuss multiple metrics at once. 

%\paragraph{Measurements' stability}

%\paragraph{Slowdown factor} 
%\begin{figure}
%    \centering
%\includegraphics[width=4cm]{mock-figs/amplif.png}
%    \caption{Amplification factor - time / amplitude}
%    \label{fig:amplif}
%\end{figure}
\section{Empirical study}
\label{sec:results}

% In this section, we present the results of our measurements study, performed using the described methodology. We first introduce our setup~\sx{sec:setup}, and present our results for the local ~\sx{res:local} and cloud~\sx{res:cloud} experiments. We then explore different ways to increase the server's resilience to CPU-based DoS attacks~\sx{res:mitigation}, and investigate whether payloads smaller than one second can be used for a successful attack~\sx{res:threshold}. Finally, we discuss the implications of our findings~\sx{sec:discussion}.

In this section, we present a measurements study performed using the described methodology. \msd{We first introduce our setup~\sx{sec:setup}, describe our validation process and results using local machines~\sx{res:local}, and present the results for the cloud experiments~\sx{res:cloud}.} We then explore different ways to increase the server's resilience to CPU-based DoS attacks~\sx{res:mitigation}, and investigate whether payloads smaller than one second can be used for a successful attack~\sx{res:threshold}. Finally, we discuss the implications of our findings~\sx{sec:discussion}.

\begin{table}[h]
  \begin{center}
  
      \begin{tabular}{c|c|c|c|c|}
      \cline{2-5}
        &\multicolumn{2}{c|}{CPU}&\multicolumn{2}{c|}{Memory}\\
      \cline{2-5}
      &Free&Pro&Free&Pro\\
      \hline
      AWS& 1 vCPU & 1 vCPU & 0.5 GB & 4 GB\\
      Azure& 1 vCPU & 1 vCPU & 1.75 GB &3.5 GB\\
      Digitalocean&1 vCPU & 1 vCPU & 512 MB & 1 GB\\
      Heroku & 1 vCPU & 1 vCPU & 512 MB & 512 MB\\
      \hline
      \end{tabular}
  \end{center}
  \caption{Hardware capabilities for considered PaaS instances.}
  \label{architecure_info}
  \vspace{-4mm}
\end{table}
\vspace{-5mm}
\subsection{Study's setup}
\label{sec:setup}

%Since all the considered cloud platforms have a framework-centered view, e.g., they offer separate developer guides for each web framework, we assume a similar view in this section. Considering that most of the time, there is a one-to-one match between the web framework and the underlying server architecture, we present our results using 

\paragraph{\textbf{Local setup}}
For our local experiments, the web servers are hosted on a MacBook Pro machine, running macOS Big Sur on Intel Core i7 6-Core CPU with 2.6 GHz and 16 GB RAM. 
% The server, attacker, and the victim are all on the same machine. 
%For all web servers. we evaluated our methodology using the default framework settings. Thus in our local setup, we did not have any load balancer and reverse proxy. 
%Different frameworks come with different default web servers. For example, Tomcat is shipped with a Tomcat web server, while Django and Express.js are shipped with their own development server. In the case of Apache, we used XAMPP server which includes an Apache server. 
%We use the default web server for each framework, whenever available: Django, Go, and Express.js are shipped with their own development server, Tomcat for Tomcat, and XAMPP for Apache.
We consider five web servers for our study: Apache, Tomcat, Node.js, the built-in Go web server, and Django's development server. We use the out-of-the-box settings for all  servers. The local experiments are intended to give us a cheap testing bed for our methodology, where we can quickly prototype a new hypothesis. We believe that these setups are only rarely used in production by inexperienced developers.
%As different frameworks are deployed using different deployment settings in practice to get the best performance, it is very difficult to create a uniform setting for all frameworks. That is why we opted to use the out-of-the-box settings for all frameworks. 

\paragraph{\textbf{Cloud setup}}
We select the following PaaS providers for our study: AWS Elastic Beanstalk, Azure Web Apps, Heroku, and  DigitalOcean App Platform. 
These four platforms are the most popular ones, controlling more than 60\% of the PaaS market share\footnote{\url{https://www.statista.com/statistics/478119/paas-vendor-market-share-ranking-worldwide/}}.
%These PaaS provide a variety of machines, depending on pricing. 
We study three configurations for each platform: a free instance, a professional instance, and two professional instances with a load balancer.
%Although it is possible to identify the exact machine architecture or system information, PaaS providers list the specification of the virtual machine on their website. 
Table~\ref{architecure_info} lists all the configurations used in the study, as detailed on the platforms' websites. As the reader may notice, there are significant differences in the hardware used by the platforms. We argue that this is beneficial for our study, since it allows us to observe the given systems under multiple, realistic deployment conditions. In the cloud setup, we study the same servers as in our local setup, except for Django, for which we study Gunicorn, which is recommended by the community.

\paragraph{\textbf{Lightweight web application}}
%In our study, we consider the following web frameworks for building a realistic web server: Express for Node.js, Django for Python, Tomcat for Java, vanilla Apache, and Go for Go. We opt for these frameworks because of their popularity among developers, and because the considered cloud providers support most of them. 
To study the given web server, we create a simple web application with three endpoints: (i) a measurement endpoint that accepts a single parameter and returns it as response, (ii) a vulnerable endpoint that simulates an algorithmic complexity vulnerability, i.e., an input-dependent slow computation, and (iii) a calibration endpoint that is used in the setup phase to fix the attack parameters. We claim that this very simple application allows us to analyze the performance of the underlying web server in the case of CPU-heavy requests.

% vulnerable endpoint
For implementing the vulnerable endpoint, we use modular exponentiation ($p^q\mod r$) to simulate the presence of a vulnerability. Modular exponentiation is very common in encryption and key generation, and it is computationally expensive. For example, in our local machine, it takes 100 ms to decrypt a 2,000 characters message that is encrypted with a 7,680-bits key, using \texttt{node-rsa}\footnote{\url{https://www.npmjs.com/package/node-rsa}} library. We fix $p$ and $r$ to two large primes and use $q$ as the payload in our requests. \msd{It is worth noting that in real systems, our specified slowdowns can be induced in many ways e.g., vulnerable regular expression engine, synchronous file or database operations. Thus, our methodology and measurement study is independent of the actual vulnerability used to trigger the slowdown.}

\begin{figure*}
    \centering
    \includegraphics[keepaspectratio=true, width=\linewidth]{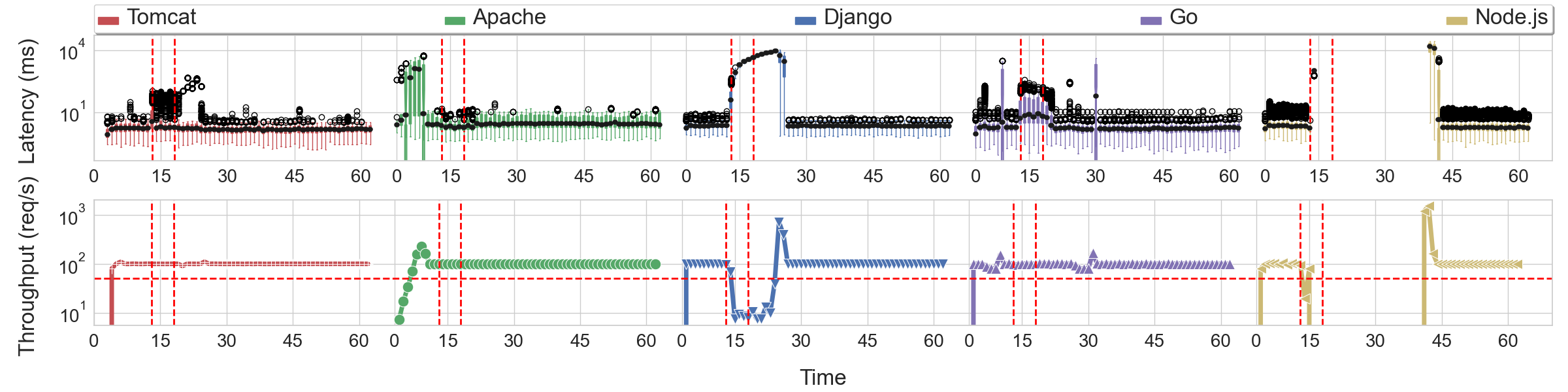}
    \caption{Throughout and latency for the considered servers in the local setup, for a $\mathcal{A}=(100, 500ms, 5s)$ attack. For each second in the time axis, we show the average throughput in that second, averaged across five runs, and a box-and-whiskers distribution of the requests' latency in that second.
    We depict the attack window between the two dashed, vertical lines and 50\% of the expected throughput with the dashed, horizontal line. We also bring the reader's attention to the  usage of logarithmic y-scales in this figure.}
    \vspace{-4mm}
    \label{fig:local_overview}
\end{figure*}

% calibration endpoint
Before the actual experiment, we perform payload calibration for each system: for a target slowdown, e.g., 100 ms, we search for the value $q$ that triggers a CPU-computation of that length. Because of the difference in underlying hardware or software, we need different $q$ values to incur the same slowdown in the same server, hosted by different providers. For example, while it takes $q=6,360,001$ to cause a one-second slowdown in AWS for Node.js, in DigitalOcean, the value is $q=17,840,001$, for the same slowdown and server. Since this process is crucial for simulating equivalent attacks against target systems, we pay special attention to ensuring the chosen $q$ value triggers the desired slowdown: (i) we perform precise measurements on the server-side, (ii) for each candidate $q$ value, we repeat the experiment ten times and compute the average computation time. \mas{To avoid the effects of caching, each request contains a random query parameter. }
\vspace{-2.5mm}
\paragraph{\textbf{Simulate attack}}
For most of the experiments, we use a five seconds attack window, we set the attack's bandwidth to 100 requests per second, and we vary the payload size from few milliseconds to one second. The measurement rate is 100 requests per second, and the measurement window is 60 seconds. We believe these are sensible parameter choices that enable us to study a short, low-bandwidth burst that could be used in on-off attacks, as described by Shan et al.~\cite{shan_tail_2017}. Moreover, it allows us to observe the server's behavior before, during, and after the attack. For all of our experiments, we simulate the attack five times and present aggregated results.

%a bandwidth-bound, small window attacker model. Our typical experiment goes as follows, the victim starts sending benign requests to the server at $t_1$ time with $100$ requests/s bandwidth. At $t_1+10s$, the attacker starts to send malicious requests to the server that continues till $t_1+15s$. For then, the server only receives requests from the victim till $t_1+60s$. For each benign request, we record the timestamps when it was sent ($T_{start}$) and when the corresponding response was received ($T_{end}$), and we compute the request processing time ($RPT = T_{end} - T_{start}$). We repeated each experiment five times to report average performance metrics ($N = 5$). To get independent observation from different experiments, we reset and restart all servers after each experiment in local setup and pause for 1 minute between two consecutive experiments. Table \ref{architecure_info} shows all the system and server information used in our experiments.%{Say how we calibrate the attack parameters}

% Describe our local setup.

% Describe the attack parameters: for Q1 and Q2 mostly 100 r/s bandwith, 5 seconds attack window, 60 seconds measurements window, different payload/slowdown levels. Try to justify why these are good attack parameters. For Q4, we use different bandwidths for the attack.

% Describe a bit the cloud setup maybe: how many instances, how did we setup each server.

% Explain our modifications to wrk2.

% \subsection{Q1: Differences in local setup}
% \vspace{-3mm}
\subsection{Local validation}
\label{res:local}
\begin{figure*}
\begin{center}
\subfigure[Attacker's gain. Each point represents the average Attacker's gain calculated over the independent simulations of the attack.]{
\includegraphics[keepaspectratio=true,angle=0,width=.28\linewidth] {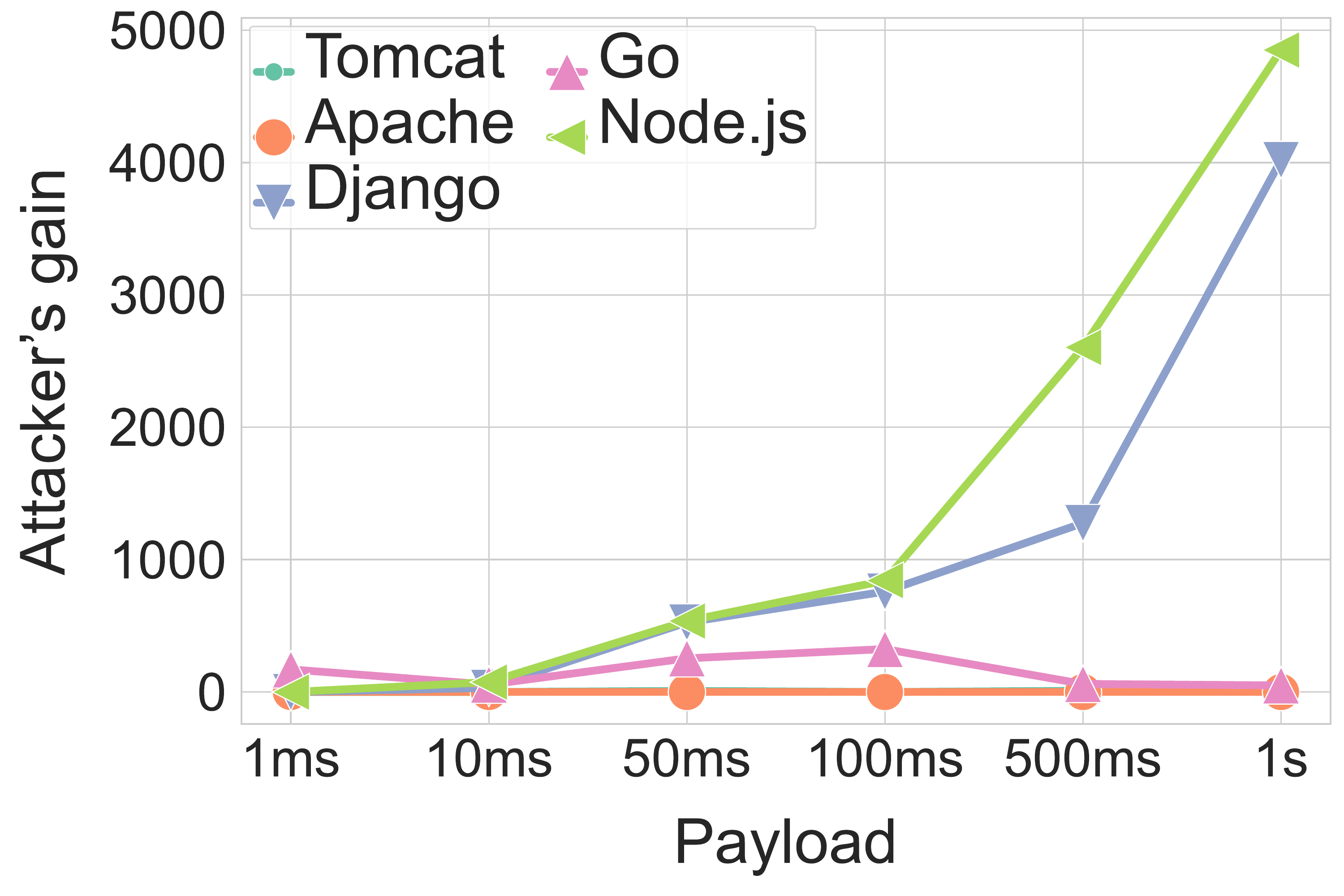}
\label{fig:gain_local}}
\hspace{0.1cm}
\subfigure[Throttled time. The boxes depict time intervals in which the throughput was between two specific fractions of the expected throughput. For example, the box for "50-26" is computed as $T_{50}-T_{25}$, using the notation introduced in Section~\ref{subsec:metrics}.]{
\includegraphics[keepaspectratio=true,angle=0,width=.36\linewidth] {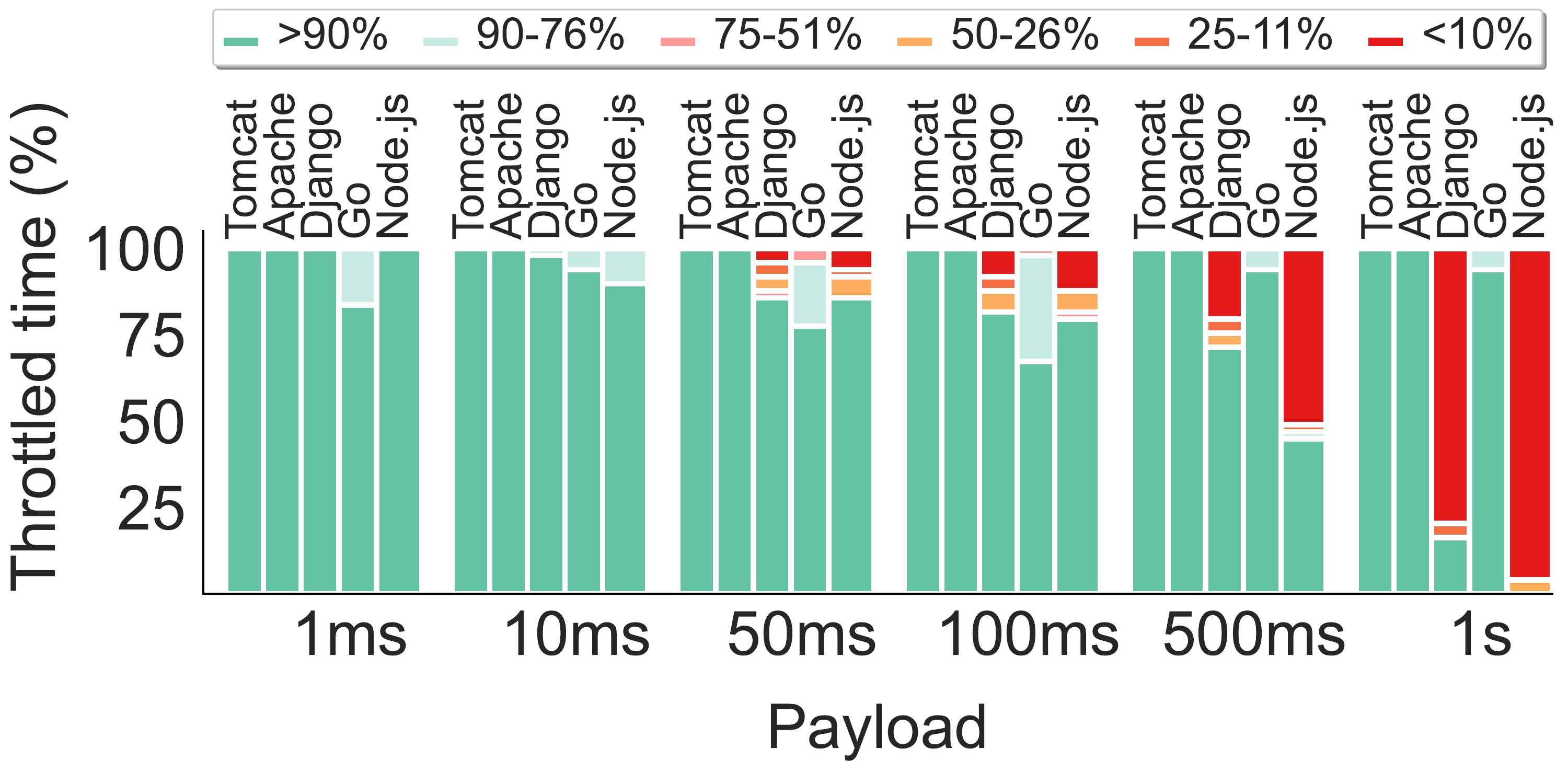}
\label{fig:throttle_time_local}}
\hspace{0.1cm}
\subfigure[Latency of requests. For each system and payload, we plot the latency of all requests observed after the attack starts. The box shows the first and third quarterly, and the whiskers show 150\% the interquartile range. ]{
\includegraphics[keepaspectratio=true,angle=0,width=.3\linewidth] {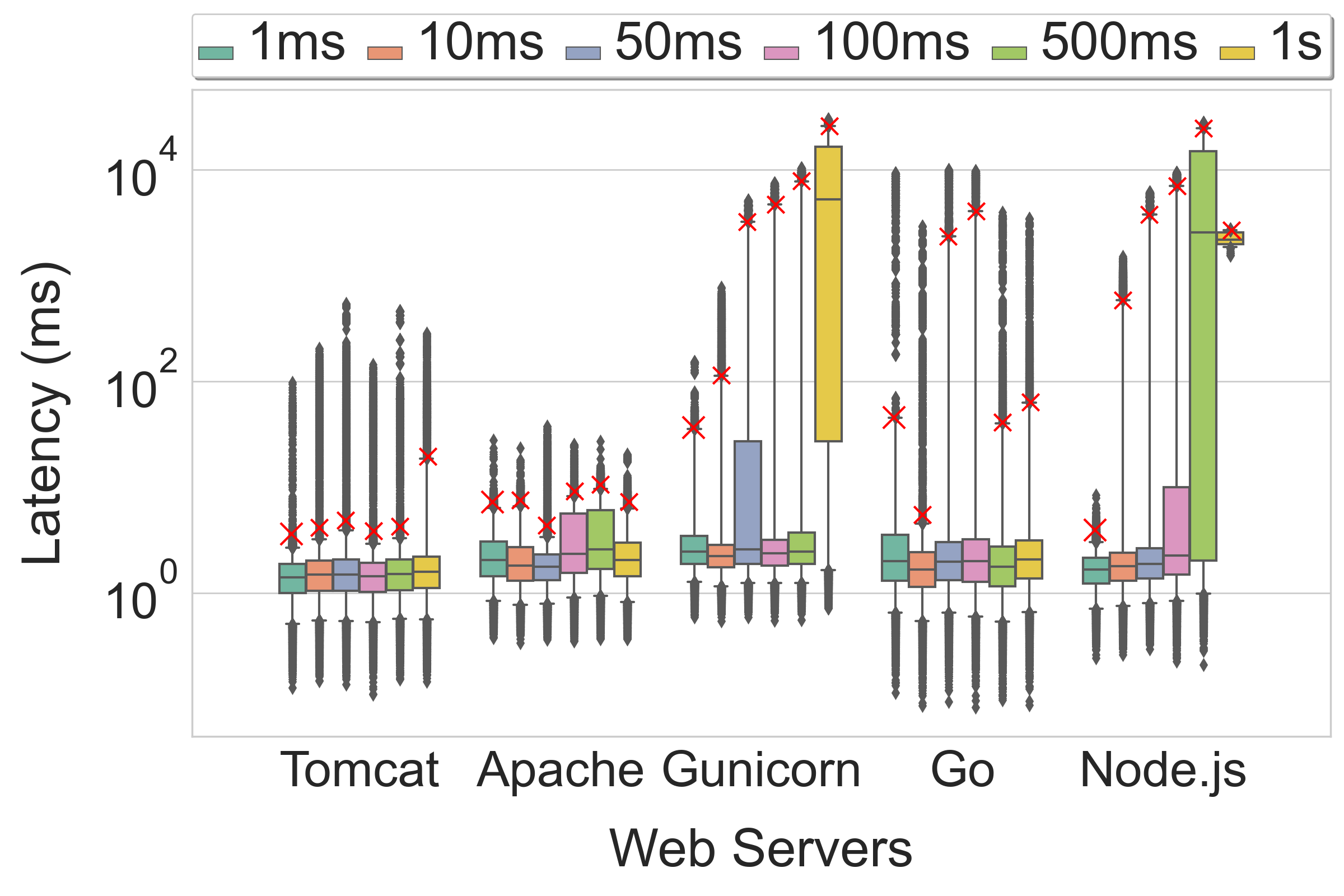}
\label{fig:request_cdf_local}}
\vspace{-2mm}
\caption{Impact of different payload sizes on Attacker’s gain~\subref{fig:gain_local}, Throttled time~\subref{fig:throttle_time_local}, and latency~\subref{fig:request_cdf_local} for the five considered web servers, in the local setup. The results correspond to five independent simulations of the attack.}
\label{fig:local_experiment}
\vspace{-4mm}
\end{center}
\end{figure*}
\msd{Before applying our methodology to PaaS platforms, we first test it in our local setup. We do this for two reasons: (1) to verify our measurements methodology in an isolated and controlled environment where the influence of external factors is low, e.g., network latency, (2) to confirm that there are indeed worth-reporting differences between web server architectures. This section explains our findings in the local setup and shows how we utilize the newly-proposed metrics for comparing the response of different servers to CPU-based DoS attacks.}

%We now proceed to show the results for our local experiments, with out-of-the-box server configurations. 
Let us first analyze the servers' response to a single attack, in our local setup: $\mathcal{A}=(100, 500ms, 5s)$. Figure~\ref{fig:local_overview} shows the throughput and latency for all the considered servers before, during, and after the attack. We depict the attack between vertical dashed lines.

% event based system worst.
The first striking observation is that for Node.js, it is possible to cause serious damage to the server's performance by using such a small payload. Even though the server starts serving requests eventually, it takes almost 25 seconds for the latency and throughput to return to the expected level. We remind the reader that our attack window is only five seconds, but its effect lingers for 25 seconds. %, hence, we say that the server's performance shows a tail effect, as defined by Shan et al.~\cite{shan_tail_2017}. 
 As seen in the figure, Django also exhibits this behavior, albeit less severe: the attack's effect is visible for slightly longer than ten seconds. However, the server continues to serve requests during this time, but at a lower rate. We believe that the difference between the two servers is because  Django uses four worker threads, while Node.js uses a single-process event loop.

% Tomcat and Go show some deterioration in performance during the attack
In contrast, the other considered systems show only minor performance deterioration: the latency increases slightly during the attack for Go and Tomcat, but the degradation is larger for Go due to a higher median and larger inter-quartile distance. Apache exhibits only a minor effect, i.e., the number of outlier requests with latency higher than average is slightly increased during the attack. However, before the actual attack, Apache goes through a warm-up phase, in which its performance is lower than that of other servers.

% Give some metrics example
Now, let us illustrate how we propose to use our metrics to depict a server's performance degradation without the need to analyze time series, like above. We compute an \mas{\textit{Attacker’s gain} \sx{attacker's_gain} } of 9, 0, 1,276, 61, and 2,603 for Tomcat, Apache, Django, Go, and Node.js, respectively.
This is consistent with the discussion above: Node.js and Django have the most serious drop in throughput, while the others barely have any reduction. To interpret a concrete \emph{Attacker's gain} value like 1,200 we must imagine the surface it represents, e.g., it can depict a drop in throughput of 40 points for 30 seconds, or a drop of 100 points (server completely inactive) for 12 seconds. We note that the theoretical maximum for \emph{Attacker's gain} in this experiment is 5,000, i.e., expected throughput is 100 requests per second, and measurement window after the attack starts is 50 seconds. 

We also measure a \emph{Throttled time} \mas{\sx{throttled_time} } at 50\% of zero, zero, ten, zero, and 26 seconds for Tomcat, Apache, Django, Go, and Node.js respectively. This, again, is aligned with the discussion above: for more than 25 seconds the throughput of Node.js is low, while for Django that is the case for 10 seconds.
In this particular experiment, the \emph{Throttled time} is more or less aligned with the drop in gain, but later in this section, we discuss cases when this is not the case. 
These two metrics do not suffice to cover all the complexities of the previous discussion of Figure~\ref{fig:local_overview}, and we need to complement our high-level view with information about request's latency, e.g., what is the typical, or maximum latency.

Let us now consider the "500ms" data points in Figure~\ref{fig:local_experiment}. They depict the same experiments as Figure~\ref{fig:local_overview}, but in a more condensed form. They show that the \emph{Attacker's gain} is the largest for Node.js and Django, as well as the \emph{Throttled time}, showing several seconds with very reduced throughput, i.e., less than 10\% of the expected throughput. The result also shows that for Django, Node.js, and Go there are several requests with more than a few seconds slowdown, but that for Node.js, this effect is the most visible, i.e., the median request takes few seconds to be served, as part of the compensation area consisting of the large spike in Figure~\ref{fig:local_overview}.
We conclude that by analyzing correlated metrics like the above, one can characterize the general performance degradation of a server without the need to look at hundreds of throughput and latency graphs.

Let us now consider the entire Figure~\ref{fig:gain_local}, showing the effect of increasing the payload size on the performance of the server. We observe a superlinear increase in gain for the event-driven architectures but no noticeable increase for the other architectures. By increasing the slowdown from 500ms to 1s, it is possible to grow the \emph{Attacker’s gain} by 342\% and 214\% in Node.js and Django, respectively. When analyzing the \emph{Throttled time}, we see again an increase in the time in which the server performs very poorly (10\% of the expected throughput). To our surprise, the transition between the state in which the server performs close to the expected throughput to the state in which it performs very poorly is very sudden. We barely observe any intermediary states for Django and Node.js. On the contrary, for Go, we see a slight degradation in performance that appears sporadically during the experiments, and no measurable change in \emph{Throttled time} for Tomcat and Apache. When analyzing the latency plot, in Node.js and Django, we observe a consistent increase in latency from small payloads to large ones. The median latency increases from 1.7ms to 
16.1s for Node.js, while for Django, it goes from 2.4ms to 5.2s from the smallest payload to the largest one. This value only increases by 12\% for Tomcat for the same payloads. \msd{The $95^{th}$ percentile latency also increases from 3.3ms and 35ms to 24s and 25s for Node.js and Django respectively, which is denoted by the red cross (x) in Figure~\ref{fig:request_cdf_local}.} Go also exhibits an increase for outlier requests: for one-second payloads, there are 252 requests that take more than one second, while for Apache and Tomcat, there are none. Such slow requests were considered security-relevant by prior work~\cite{shan_tail_2017}.

On the one hand, our results so far confirm the assertion made by previous work: in their out-of-the-box setup, event-based systems are more vulnerable to CPU-based DoS attacks than traditional architectures. On the other hand, payloads as low as 100ms can be used to make a server inactive for several seconds, disproving the hypothesis that there is a special threshold below which CPU-heavy payloads are inoffensive. Now let us proceed to study more realistic setups.

%Nodejs also suffers severely against DoS attacks in Digitalocean app platforms. Although attackers gain is higher for Node.js in Digitalocean than Heroku, the server has less downtime in the former platform. This indicates even though Node.js shows more resiliency against the attack by holding the throughput above 10\% level, the effect of the attack continues through larger portion of time. Which in return, contributes to the higher latency distribution in figure \ref{fig:request_cdf_digitalocean}. 

\vspace{-2mm}
\subsection{Q1: Differences in the cloud}
\label{res:cloud}

\begin{table*}[h]
  \begin{center}
  \resizebox{\textwidth}{!}{
      \begin{tabular}{ p{0.11\linewidth} | p{0.11\linewidth}| p{0.11\linewidth} | p{0.11\linewidth}| p{0.11\linewidth} | p{0.11\linewidth} | p{0.11\linewidth} | p{0.11\linewidth}| p{0.11\linewidth}|}
        \cline{2-9}
        \multicolumn{1}{c|}{} & \multicolumn{2}{c|}{AWS} & \multicolumn{2}{c|}{Azure} & \multicolumn{2}{c|}{Digitalocean} & \multicolumn{2}{c|}{Heroku}\\ %& \multicolumn{3}{c|}{Local}\\
        \cline{2-9}
        &Throughput&Latency&Throughput&Latency&Throughput&Latency&Throughput&Latency\\
        \cline{2-9}
        Tomcat &  0.00 ±  0.00 &  0.11 ±  0.05 &  0.01 ±  0.04 &  0.07 ±  0.11 &  0.01 ±  0.04 &  0.23 ±  0.09 &  0.03 ±  0.04 &  0.06 ±  0.05\\
        \hline
        Apache &  0.01 ±  0.01 &  0.30 ±  0.35 &  0.04 ±  0.18 &  0.06 ±  0.03 &  0.02 ±  0.02 &  0.48 ±  0.08 &  0.04 ±  0.05 &  0.04 ±  0.03\\
        \hline
        Gunicorn &  0.01 ±  0.01 &  0.22 ±  0.14 &  0.01 ±  0.02 &  0.15 ±  0.09 &  0.00 ±  0.00 &  0.47 ±  0.04 &  0.01 ±  0.01 &  0.09 ±  0.08\\
        \hline
        Go &  0.02 ±  0.05 &  0.13 ±  0.06 &  0.00 ±  0.00 &  0.09 ±  0.10 &  0.02 ±  0.05 &  0.22 ±  0.10 &  0.03 ±  0.05 &  0.05 ±  0.06\\
        \hline
        Node.js &  0.00 ±  0.00 &  0.15 ±  0.12 &  0.07 ±  0.26 &  0.14 ±  0.17 &  0.00 ±  0.01 &  0.31 ±  0.13 &  0.03 ±  0.03 &  0.15 ±  0.30\\

        \hline
        % \bottomrule
      \end{tabular}
    }
  \end{center}
  \caption{\msd{Root mean square percentage error (RMSPE) of throughput and latency, in stable non-attack conditions, over five different runs. The true values for this calculation are the expected throughput (100 requests/second) and the average latency of a request in the considered period.}}
  \vspace{-5mm}
  \label{rmspe}
%   \vspace{-3mm}
\end{table*}

Figures~\ref{fig:aws_experiment}-\ref{fig:heroku_experiment} show our experimental results in the four considered PaaS platforms. We use one paid, professional instance for all the experiments. \msd{To increase confidence in our measurements, we start by presenting the root mean square percentage error (RMSPE) values of both throughput and latency. Table~\ref{rmspe} shows the RMSPE values for all considered web servers on all platforms, in stable non-attack conditions. In our attack scenario, the attack starts after 10s, and sometimes it takes a few seconds for the server to warm up. Therefore, we consider the 5\textsuperscript{th}-10\textsuperscript{th} seconds as the stable period and calculate RMSPE over this period. The first column of the table shows the RMSPE values for throughput for five different runs. We use the expected throughput as the true value for the RMSPE calculation. The data shows that throughput remains stable during non-attack conditions as the error percentage is $\leq10\%$ in all cases, except for one where it is close to $30\%$. The second column shows the error percentage for latency during the same runs. Unlike throughput, latency depends on multiple components and can vary highly depending on the conditions. Our data shows the latency error was $\leq20\%$ in most cases except for a few outliers. We believe that these values are small enough to allow us to draw important conclusions about the analysed servers.

\begin{figure*} [p]
\begin{center}
\subfigure[Attacker's gain.]{
\includegraphics[keepaspectratio=true,angle=0,width=.28\linewidth] {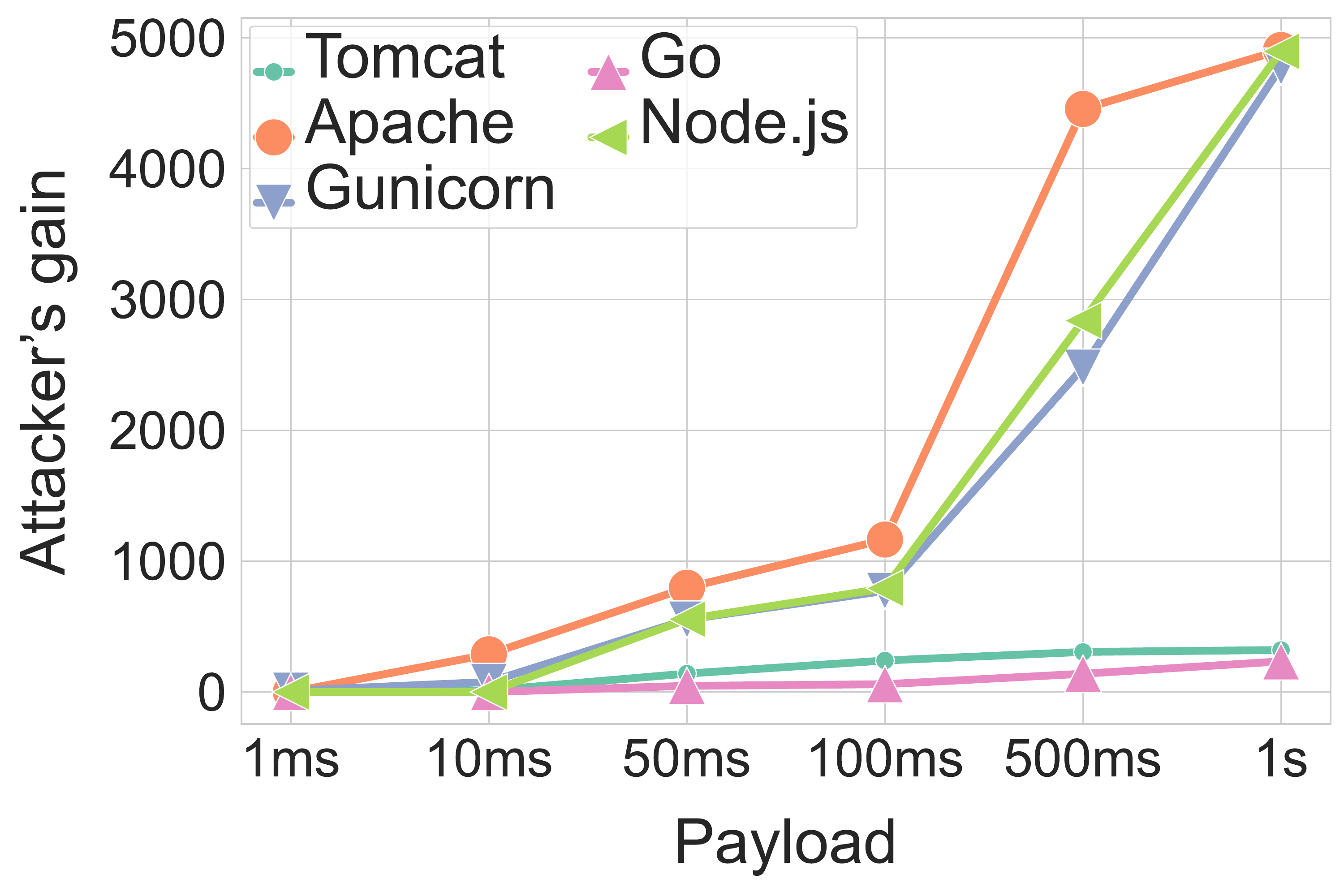}
\label{fig:gain_aws}}
\subfigure[Throttled time.]{
\includegraphics[keepaspectratio=true,angle=0,width=.38\linewidth] {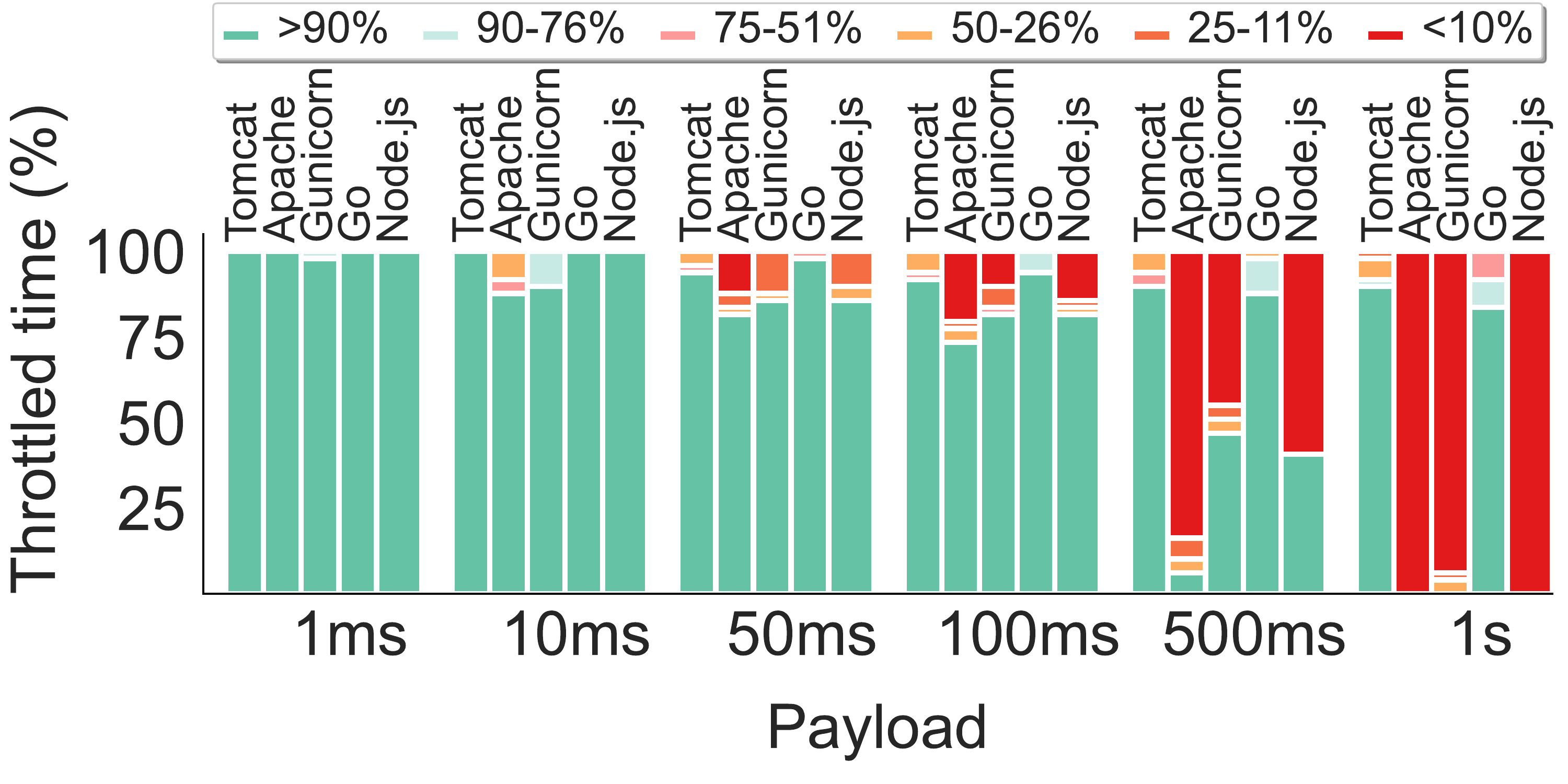}
\label{fig:throttle_time_aws}}
\subfigure[Latency of requests.]{
\includegraphics[keepaspectratio=true,angle=0,width=.3\linewidth] {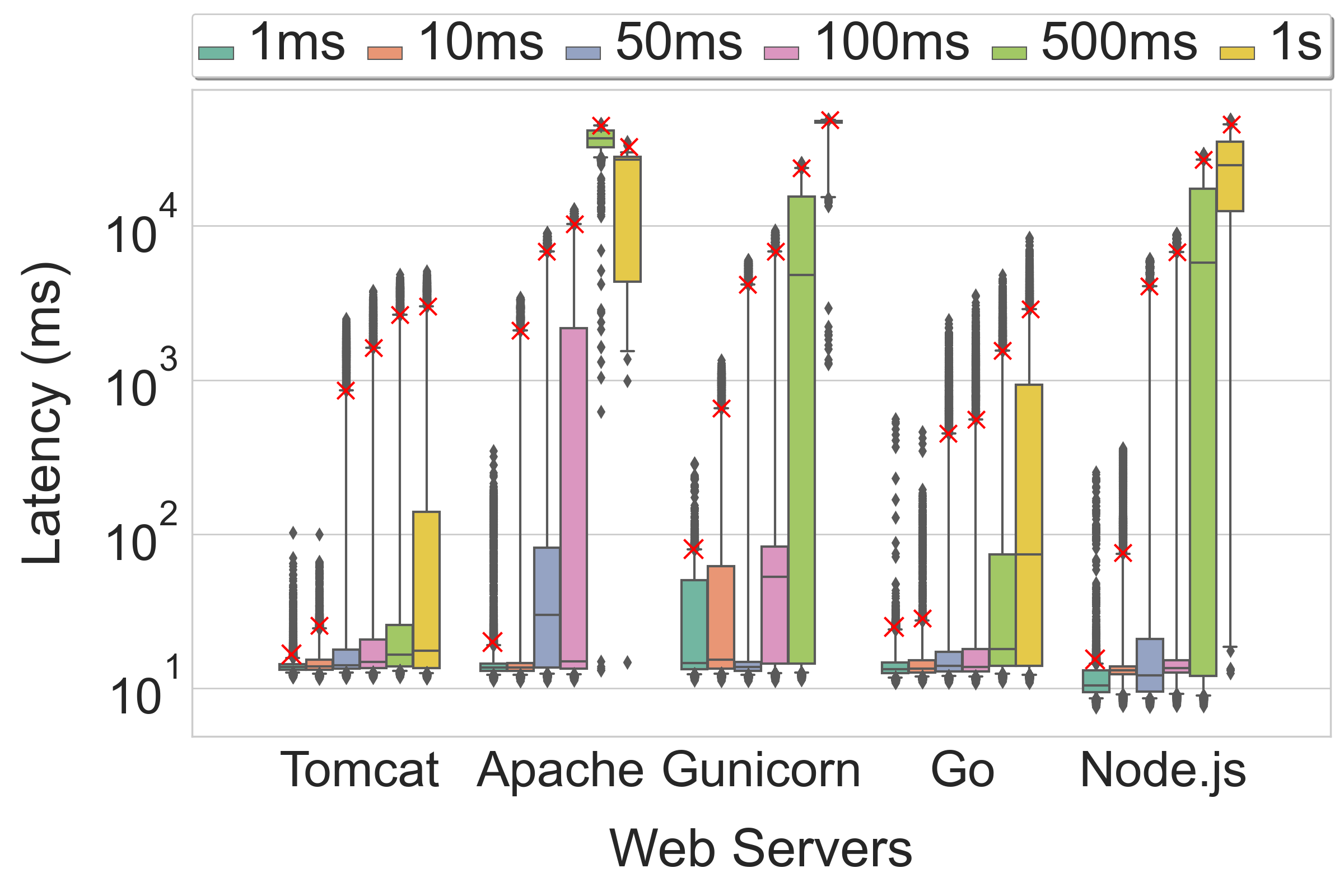}
\label{fig:request_cdf_aws}}
\caption{AWS experiment. We use the same notation as in Figure~\ref{fig:local_experiment}.}
\label{fig:aws_experiment}
\subfigure[Attacker's gain.]{
\includegraphics[keepaspectratio=true,angle=0,width=.28\linewidth] {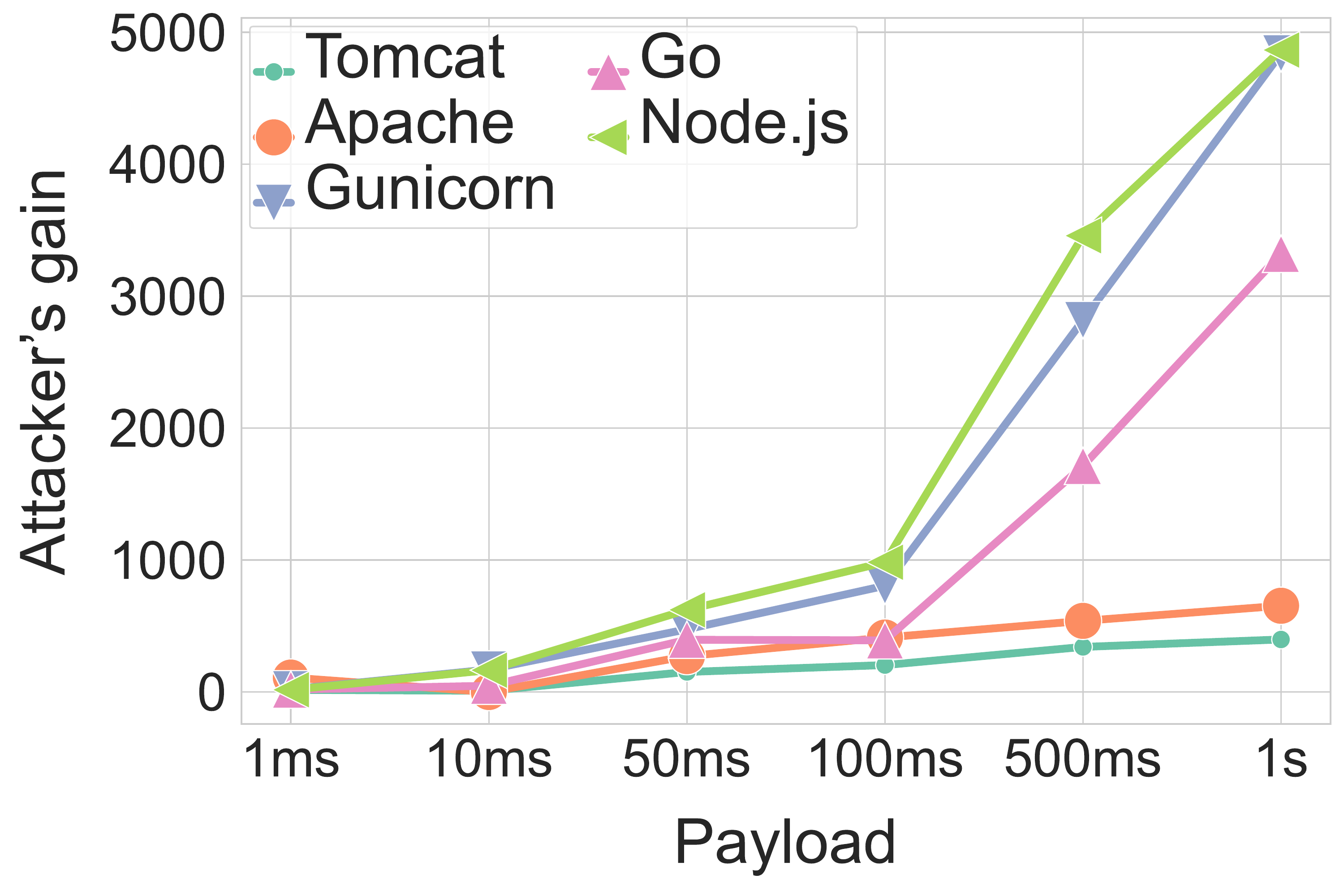}
\label{fig:gain_azure}}
\subfigure[Throttled time.]{
\includegraphics[keepaspectratio=true,angle=0,width=.38\linewidth] {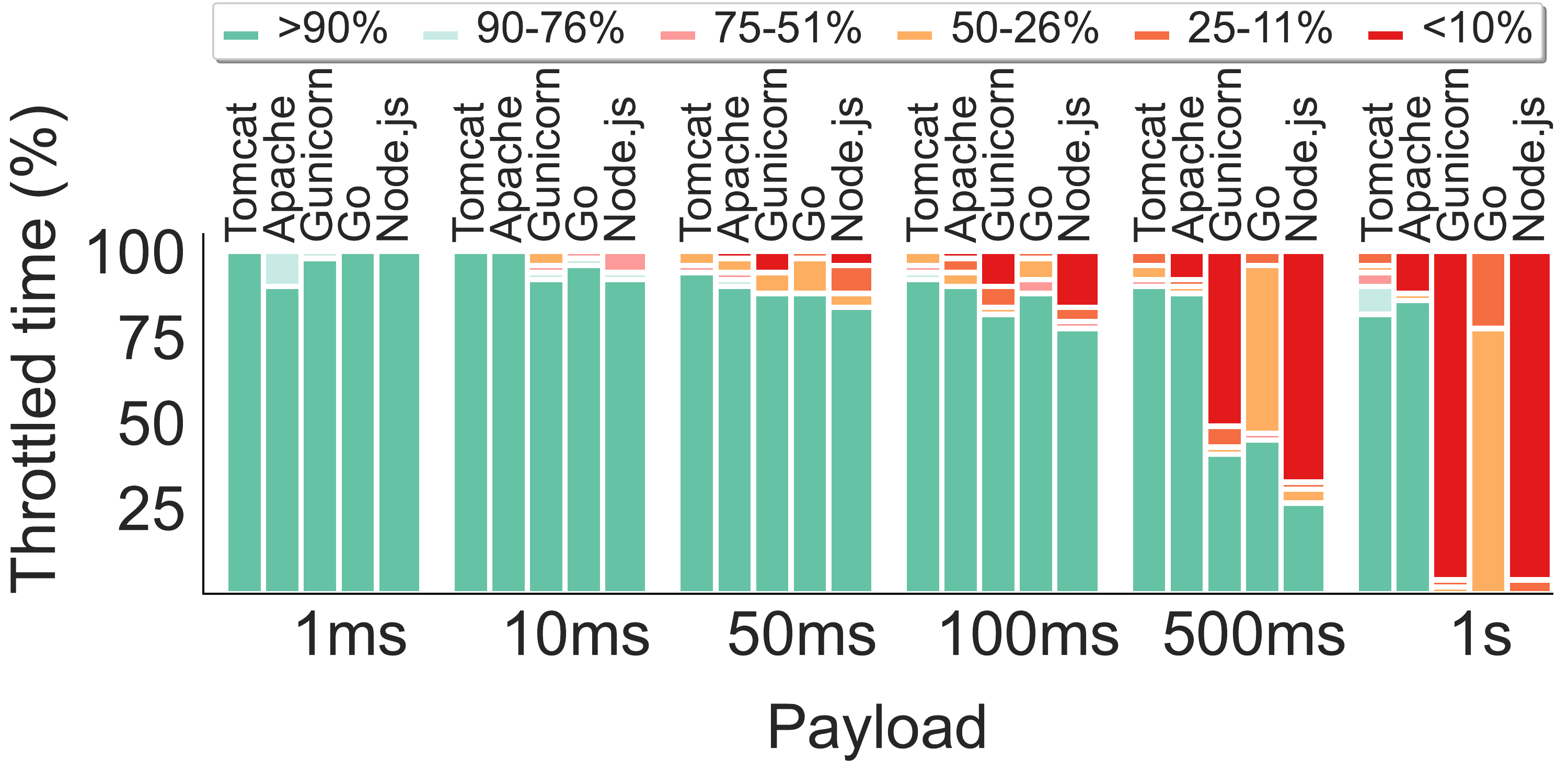}
\label{fig:throttle_time_azure}}
\subfigure[Latency of requests.]{
\includegraphics[keepaspectratio=true,angle=0,width=.3\linewidth] {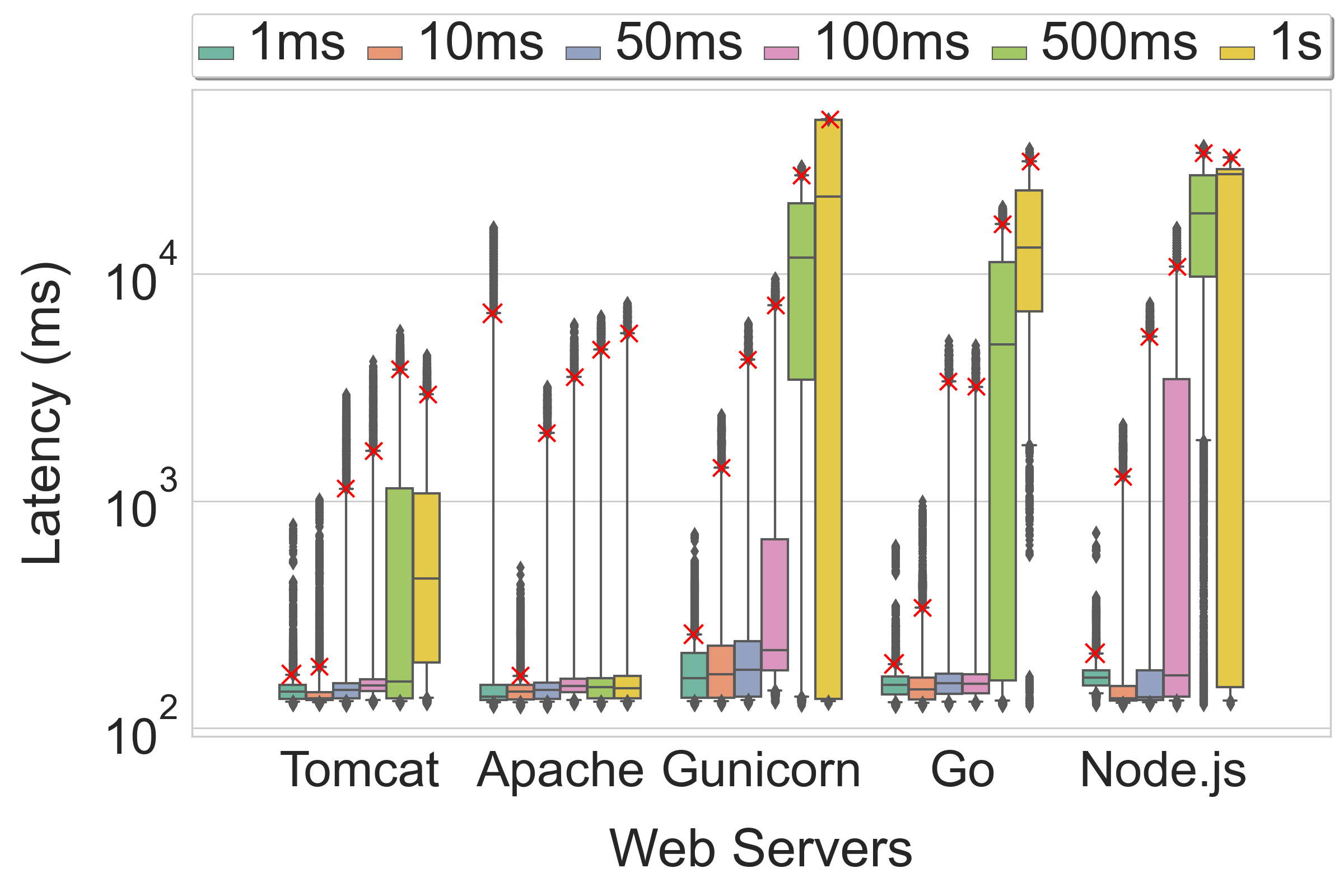}
\label{fig:request_cdf_azure}}
\caption{Azure experiment. We use the same notation as in Figure~\ref{fig:local_experiment}.}
\label{fig:azure_experiment}
\subfigure[Attacker's gain.]{
\includegraphics[keepaspectratio=true,angle=0,width=.28\linewidth] {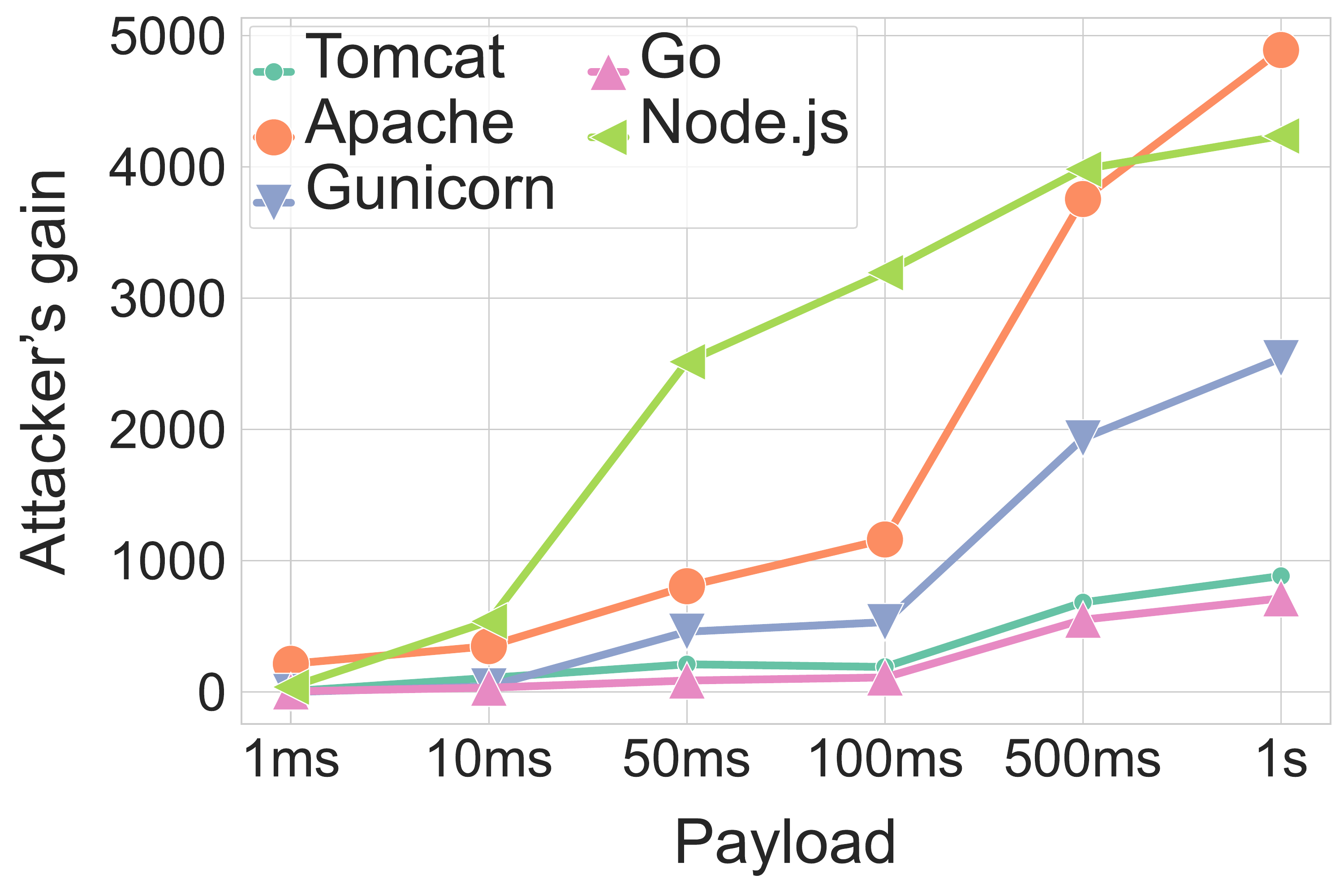}
\label{fig:gain_digitalocean}}
%\hspace{3mm}
\subfigure[Throttled time.]{
\includegraphics[keepaspectratio=true,angle=0,width=.38\linewidth] {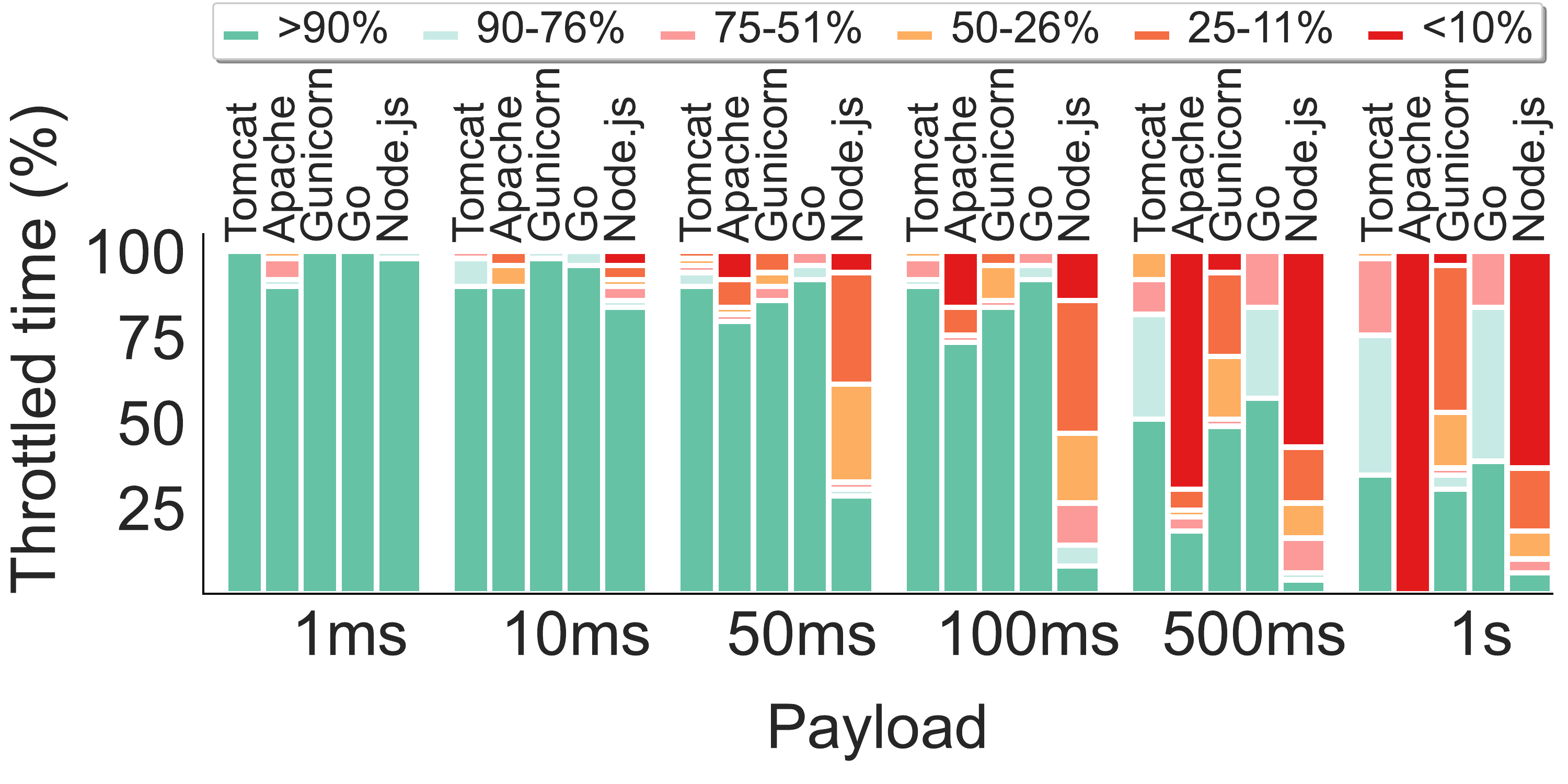}
\label{fig:throttle_time_digitalocean}}
%\hspace{3mm}
\subfigure[Latency of requests.]{
\includegraphics[keepaspectratio=true,angle=0,width=.3\linewidth] {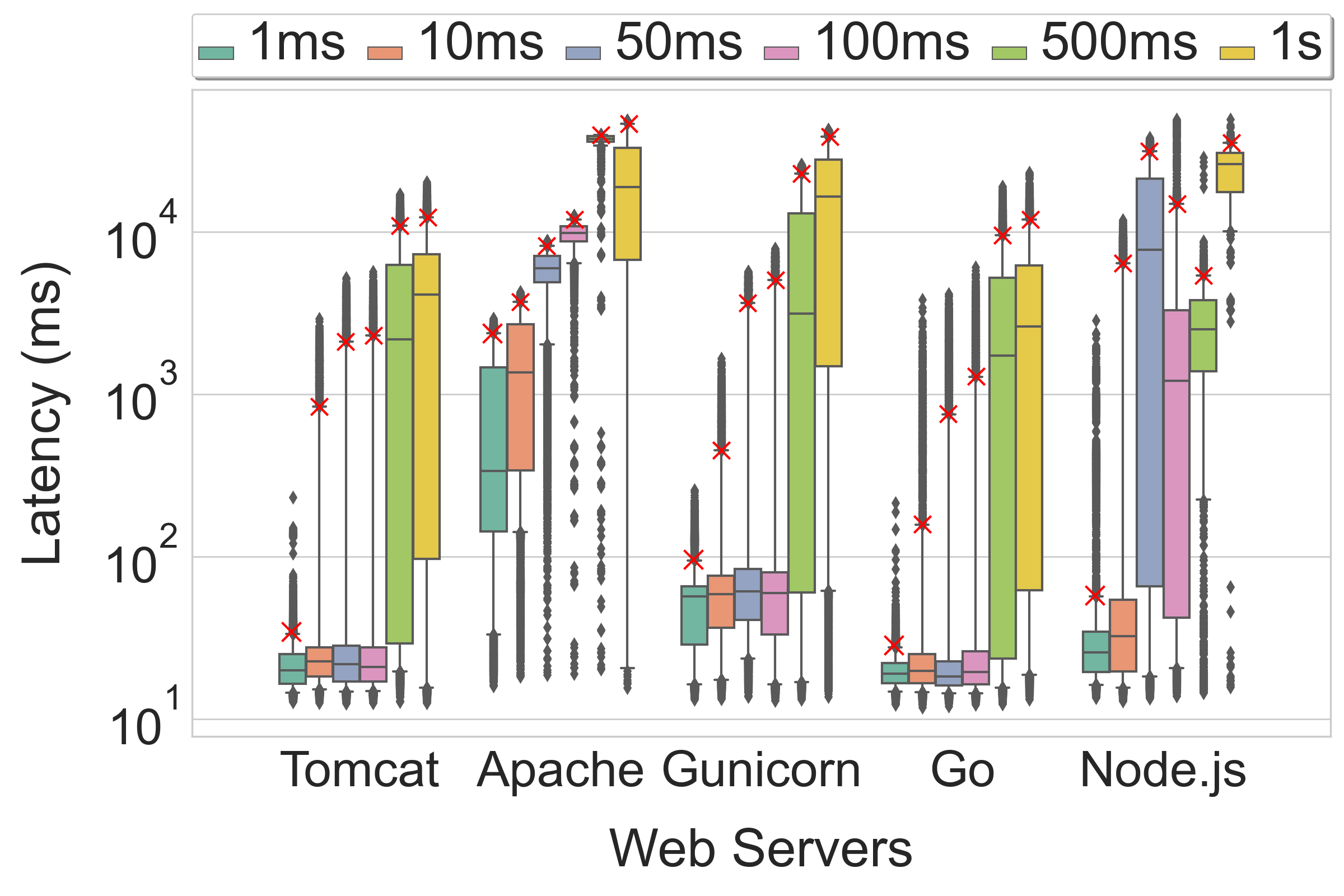}
\label{fig:request_cdf_digitalocean}}
\caption{DigitalOcean experiment. We use the same notation as in Figure~\ref{fig:local_experiment}.}
\label{fig:digitalocean_experiment}
\subfigure[Attacker's gain]{
\includegraphics[keepaspectratio=true,angle=0,width=.28\linewidth] {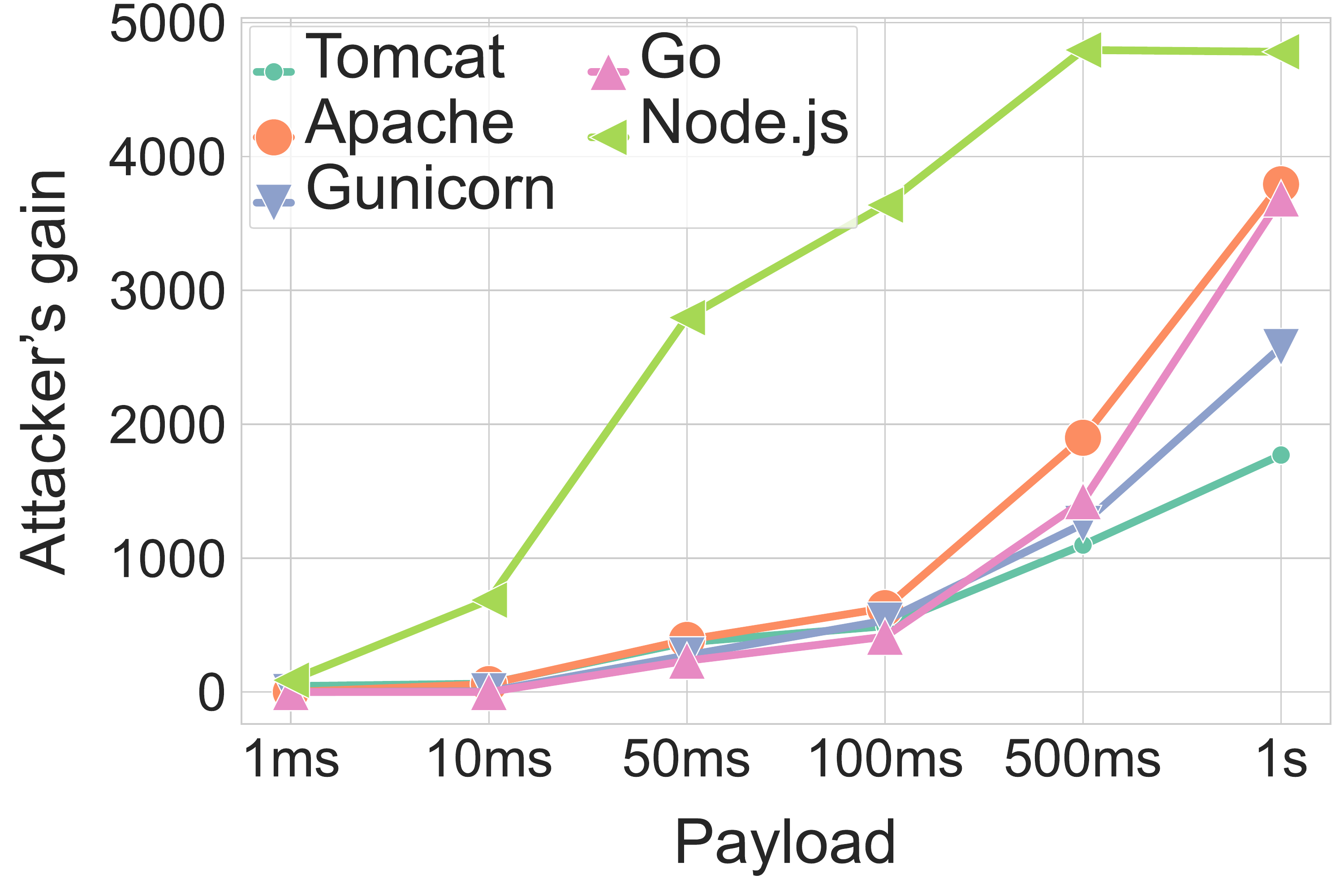}
\label{fig:gain_heroku}}
\subfigure[Throttled time.]{
\includegraphics[keepaspectratio=true,angle=0,width=.38\linewidth] {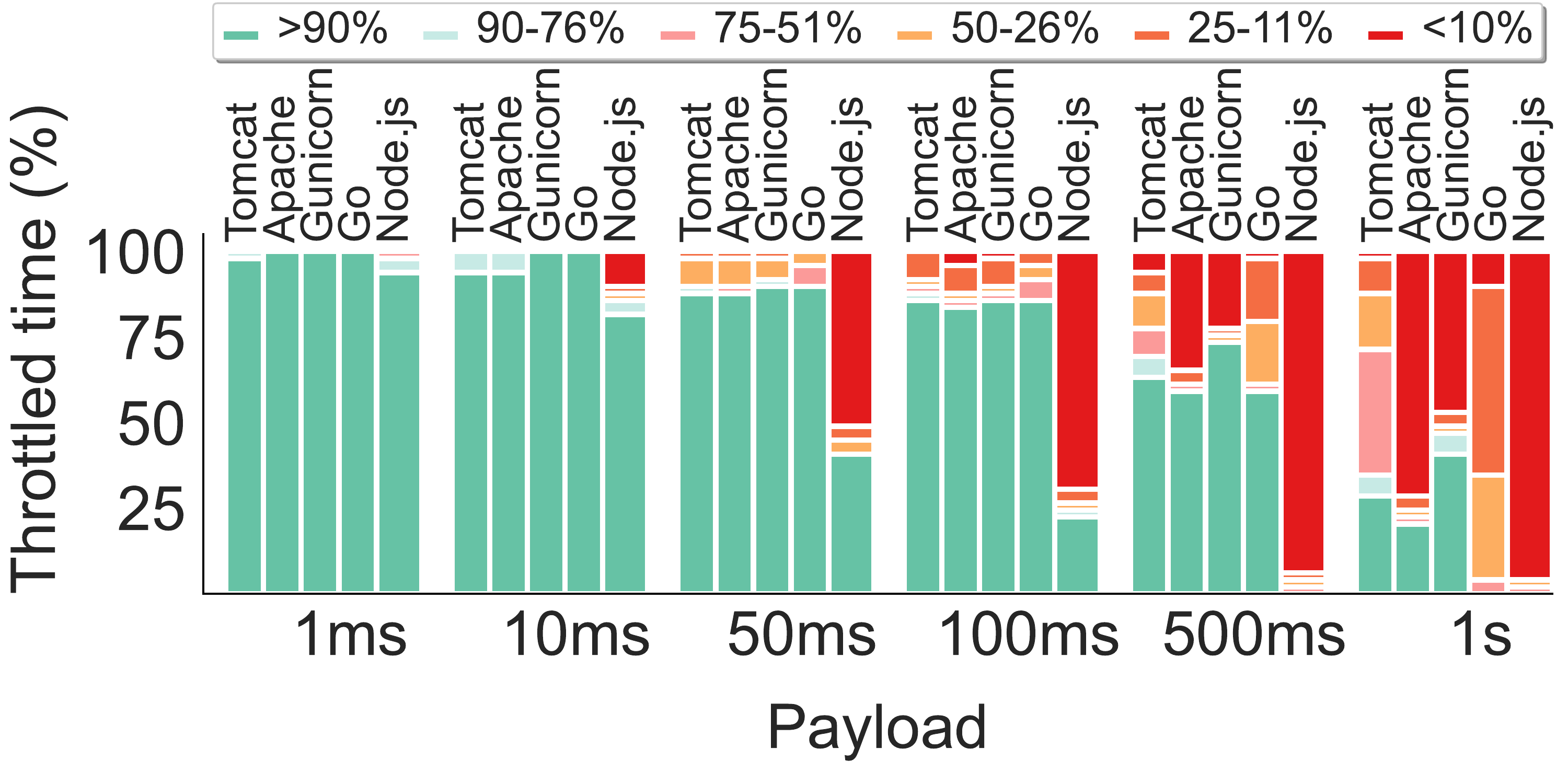}
\label{fig:throttle_time_heroku}}
\subfigure[Latency of requests.]{
\includegraphics[keepaspectratio=true,angle=0,width=.3\linewidth] {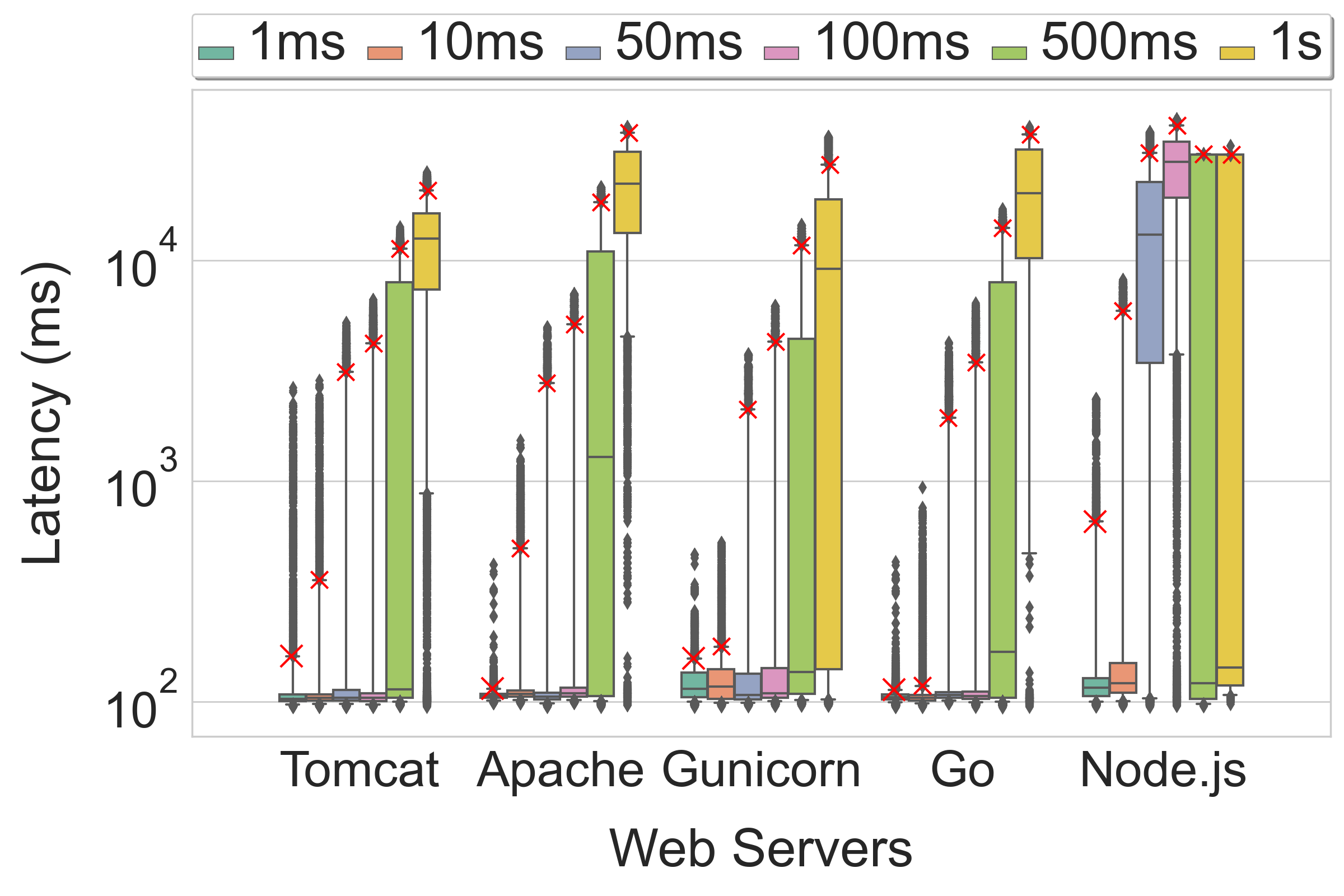}
\label{fig:request_cdf_heroku}}
\caption{Heroku experiment. We use the same notation as in Figure~\ref{fig:local_experiment}.}
\label{fig:heroku_experiment}
\end{center}
\end{figure*}

It is worth noting that, the RMSPE values for the entire first 10 seconds do not deviate much from the previous one. However, we notice that some Apache instances have high latency at the beginning of this time interval, and we hypothesize that it is because they only initialize the thread pool once requests start arriving. }

We \msd{can also confirm this} by observing that the gain for the smallest considered payload (1ms) is always very close to zero, showing that when the attacker does not possess powerful enough payloads, the server approaches DoS noninterference, i.e., the observer's throughput is almost perfectly matching the expected throughput. These data points \msd{also} serve as a quick validation for our measurements, showing that the throughput of the observer is the expected one, in non-attack conditions. 
% Lack of aggressive rate limiting or IDS
Related to this, we do not observe any form of rate-limiting enforced by the cloud platform. During our measurements, we could send several hundred requests per second without being restricted in any way, e.g., by an intrusion detection system.

% Azure is the one that closest matches our local results
The Azure results (Figure~\ref{fig:azure_experiment}) appear to best match our local ones: the performance degradation is most serious for Gunicorn and Node.js, while for Tomcat and Apache is very low. Contrary to our local setup, the attack's impact on latency, for Tomcat is more significant for large payloads, causing several requests to be served after one second. Moreover, Go shows much more serious performance degradation: \emph{Attacker's gain} is much higher than in our local experiment, and the throughput is less than 50\% of the expected one, for most of the time after the attack started.

% \begin{table*}[h]
%   \begin{center}
%   \resizebox{\textwidth}{!}{
%       \begin{tabular}{ p{0.11\linewidth} | p{0.11\linewidth}| p{0.11\linewidth} | p{0.11\linewidth}| p{0.11\linewidth} | p{0.11\linewidth} | p{0.11\linewidth} | p{0.11\linewidth}| p{0.11\linewidth}|}
%         \cline{2-9}
%         \multicolumn{1}{c|}{} & \multicolumn{2}{c|}{AWS} & \multicolumn{2}{c|}{Azure} & \multicolumn{2}{c|}{Digitalocean} & \multicolumn{2}{c|}{Heroku}\\ %& \multicolumn{3}{c|}{Local}\\
%         \cline{2-9}
%         &Throughput&Latency&Throughput&Latency&Throughput&Latency&Throughput&Latency\\
%         \cline{2-9}
%         Tomcat &0.00 ±0.00 &15.27 ±2.98 &1.16 ±3.67 &146.09 ±16.01 &1.24 ±4.28 &22.85 ±4.43 &2.63 ±3.74 &104.25 ±4.01\\
%         \hline
%         Apache &0.70 ±0.57 &24.06 ±22.34 &3.66 ±17.90 &145.88 ±9.86 &1.54 ±1.94 &190.94 ±65.70 &3.61 ±5.21 &106.93 ±3.09\\
%         \hline
%         Gunicorn &0.76 ±1.14 &28.44 ±20.55 &0.58 ±1.54 &178.20 ±34.78 &0.09 ±0.33 &53.42 ±8.58 &1.02 ±1.08 &117.76 ±17.53\\
%         \hline
%         Go &2.25 ±5.44 &15.25 ±3.10 &0.19 ±0.33 &152.33 ±21.05 &1.62 ±4.62 &21.97 ±3.86 &2.86 ±4.88 &105.16 ±3.23\\
%         \hline
%         Node.js &0.00 ±0.00 &12.32 ±3.13 &7.09 ±25.76 &195.67 ±330.02 &0.31 ±0.62 &27.51 ±9.22 &2.80 ±2.60 &116.72 ±17.93\\
%         \hline
%         % \bottomrule
%       \end{tabular}
%     }
%   \end{center}
%   \caption{Machines' capabilities, for the considered cloud setups.}
%   \label{architecure_info}
% \end{table*}

% AWS is the second, but Apache is weird
The AWS results (Figure~\ref{fig:aws_experiment}) are also mostly consistent with our local ones. However, Apache exhibits the worst performance degradation on this platform, even worse than that of event-driven systems. We hypothesize that AWS could use Apache in one process per client mode. It is also worth mentioning that the average latency for 1s payloads in Go is higher than the one for Gunicorn with a 50ms attack, even though the gain is higher for the latter. That is because Go tends to attenuate the attack over time, without a dramatic momentary reduction in throughput. In contrast, for Gunicorn, immediately after the attack starts, the throughput falls below 10\%, a trend also seen in our local experiments. %in Figure~\ref{fig:local_experiment}.

% In DO we start seeing degradation of performance for the other systems
Figure~\ref{fig:digitalocean_experiment} shows our result in DigitalOcean. All servers are much more impacted than in all the previous experiments: (i) for large payloads, we see a significant \emph{Attacker's gain} for all systems, (ii) the throughput for one-second payloads is almost always below 90\% of the expected throughput, and (iii) the median latency for all servers is larger than one second. Moreover, Node.js and Gunicorn show unusual behavior that we do not see in any other setup: the degradation of throughput is gradual, more similar to what we would expect from a thread-based system. In Appendix~\ref{apx:examples}, we show that the actual throughput after the attack looks very peculiar: it is most of the time zero, with large spikes from time to time. We also observe these spikes for other systems deployed on DigitalOcean. We hypothesize that they are caused by the load balancer forwarding requests out-of-order to the server, i.e., measurement requests are interleaved with attack requests. Further experiments are needed to confirm this hypothesis.

% In Heroku, the differences are the least pronounced 
Figure~\ref{fig:heroku_experiment} presents our experimental results in Heroku for the selected web servers. Node.js suffers significant degradation due to the attack: the Throttled time figure indicates that even for small payloads like 50 or 100ms, the server becomes almost completely unresponsive for more than 70\% of the time. Even though the attacker gain is comparable in Gunicorn, Go, and Apache, all three systems react differently to the DoS attack. While the throughput of Gunicorn degrades drastically as soon as the attack starts, it stabilizes the throughput to the expected level not long after the attack stops, i.e., in the \emph{Throttled time} graphs, one can see that Gunicorn recovers the fastest from the attack, serving for the longest time at >90\% of expected throughput. Contrarily, Go barely goes under 10\% of the expected throughput, even during attacks with large payloads, but for most of the time, it serves requests at less than half of the expected throughput, and the latency of the requests is, overall, higher.

% We hypothesize that it has to do with the hardware: traditional systems use resources more efficiently, during the attack
Unlike AWS and Azure, our results in Heroku and DigitalOcean show that Tomcat servers are not fully resilient against CPU-based DoS attacks with large payloads like 500ms or 1s. Figure \ref{fig:request_cdf_heroku} and \ref{fig:request_cdf_digitalocean} show that the median latency increases from 102 ms to 12.5s in Heroku, and 19 ms to 4.1s in Digitalocean. We believe this is due to the difference in hardware provided by these platforms. While AWS and Azure provide four and 3.5 GB memory respectively, Heroku and Digitalocean provide less than 1 GB. As Tomcat spawns a new thread for each request, the memory footprint is larger for Tomcat than for other systems. Thus, this server does not perform optimally in platforms with low computational resources. %In general, our results support the hypothesis that when the underlying machine has modest hardware, the difference between the servers is smaller than in cases of high performance machines. 
In the following section, we discuss the idea of increasing the hardware capabilities as a way to mitigate CPU-based DoS.

% Gradual vs. sudden degradation of performance
We observe that in most of the cloud experiments Gunicorn and Node.js confirm our local results and show a very sudden drop in performance after the attack, followed by a very sudden recovery. This is because the process(es) that serve attacker payloads quickly get clogged, unable to serve any additional incoming requests.
On the contrary, thread-based systems show a much more gradual degradation of performance caused by the attacker slowly depleting the CPU resources, one thread at a time.

% long-tail latency problem of the target web application (e.g., 95th percentile response time > 1 second)
We also notice that for 50 ms payloads and higher, virtually all the servers in all the configurations exhibit a long-tail latency. That is, several measurement requests are served in more than a second. There is an extensive body of work studying this effect~\cite{PuKLZLPWJXMWJKS17,shan_tail_2017,TennagePJJ19,ZhangSWLYW19,WangLKZP17,BergerBZ0H18}, but to the best of our knowledge, we are the first to show that a low-bandwidth DoS attack with relatively small payloads produces this effect against web applications deployed in the cloud, consistently.

\vspace{-2mm}
\subsection{Q2: Increase resilience}
\label{res:mitigation}

Our previous results show that the resiliency of a server architecture depends highly on the system configuration of the platform. For example, Tomcat excels highly in our local setup, AWS, and Azure, while it shows a slight performance degradation in Heroku and DigitalOcean against large payloads. Therefore, we study if changing the instance type, the number of instances, or other configurations affect the server's resilience to DoS attacks. We consider three configurations for each platform: an entry-level instance, a professional instance, and two professional instances.

\begin{figure*}
    \centering
    \includegraphics[keepaspectratio=true,width=\linewidth]{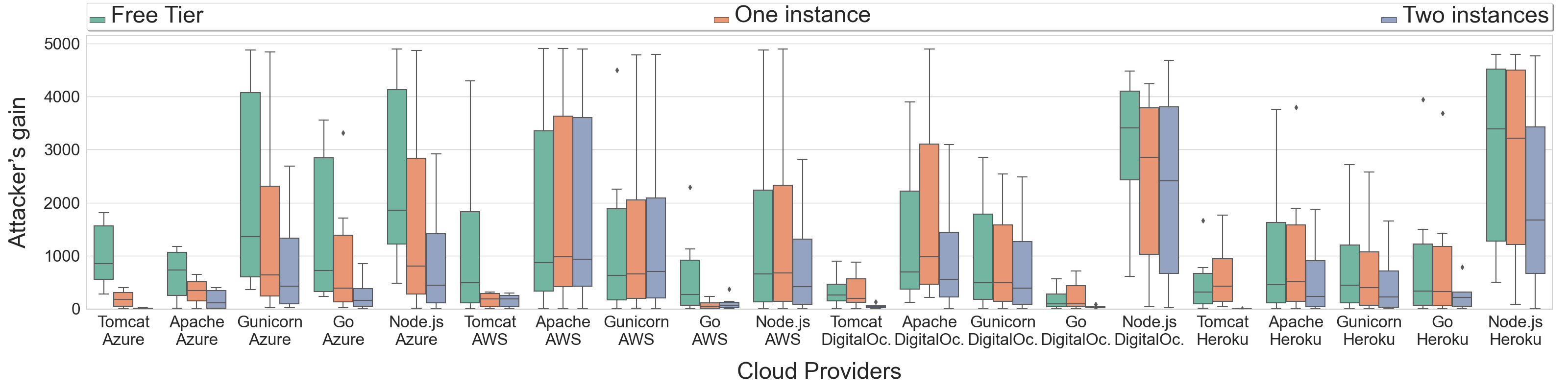}
    \caption{Impact of using free, professional, and multiple machines on Attacker's gain, in different cloud platforms.}
    \vspace{-3mm}
    \label{fig:free_vs_pro_vs_multi_machines}
\end{figure*}

Figure \ref{fig:free_vs_pro_vs_multi_machines} shows the average \emph{Attacker's gain} among these three configurations in the considered platforms. For each configuration, we repeat our experiments with increasing payloads and compute the average \emph{Attacker's gain}. For most servers, increasing the capabilities of the instance, i.e., going from entry-level to professional, result in a reduction in \emph{Attacker's gain}. A noticeable exception is Heroku, where the switch between the two types of machine is not reflected in better hardware, but in guarantees about the machine's availability. Similarly, adding more instances in parallel does appear to help as well, in most of the cases. While this naive mitigation strategy provides a clear benefit, we believe developers are not willing to pay the extra (monthly) cost to protect against threats that may never happen.

\begin{figure*} [ht]
\begin{center}
\subfigure[Attacker's gain for Tomcat vs. Node.js with multiple workers. For each PaaS platform, we consider number of workers  between one and 20, in increments of five.]{
\includegraphics[keepaspectratio=true,angle=0,width=.28\linewidth] {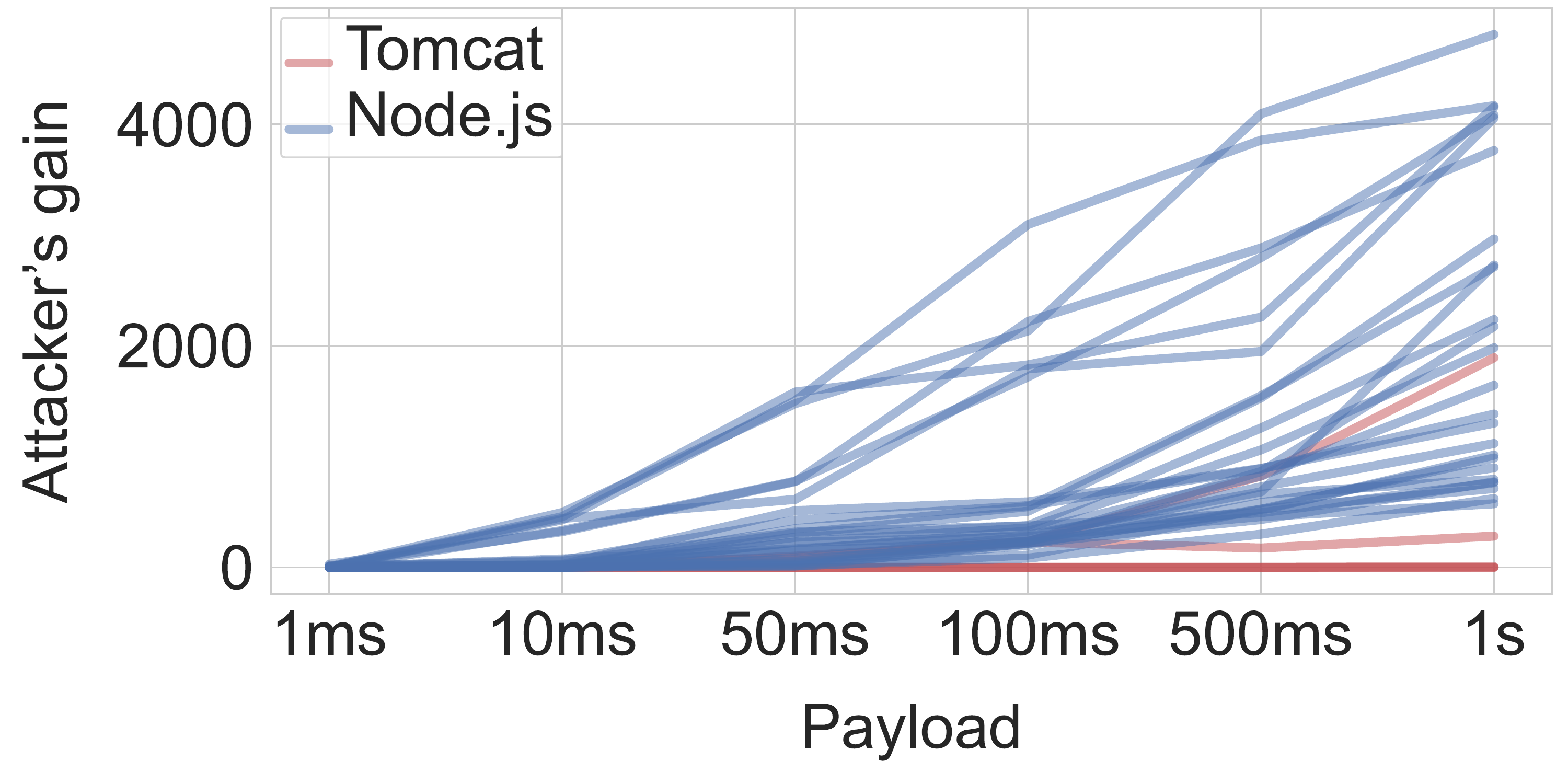}
\label{fig:workers_vs_gain}}
\hspace{0.1cm}
\subfigure[Latency in non-attack conditions. Most PaaS platforms do not allow more than 50 preforked workers, and none of them allow more than 100.]{
\includegraphics[keepaspectratio=true,angle=0,width=.3\linewidth] {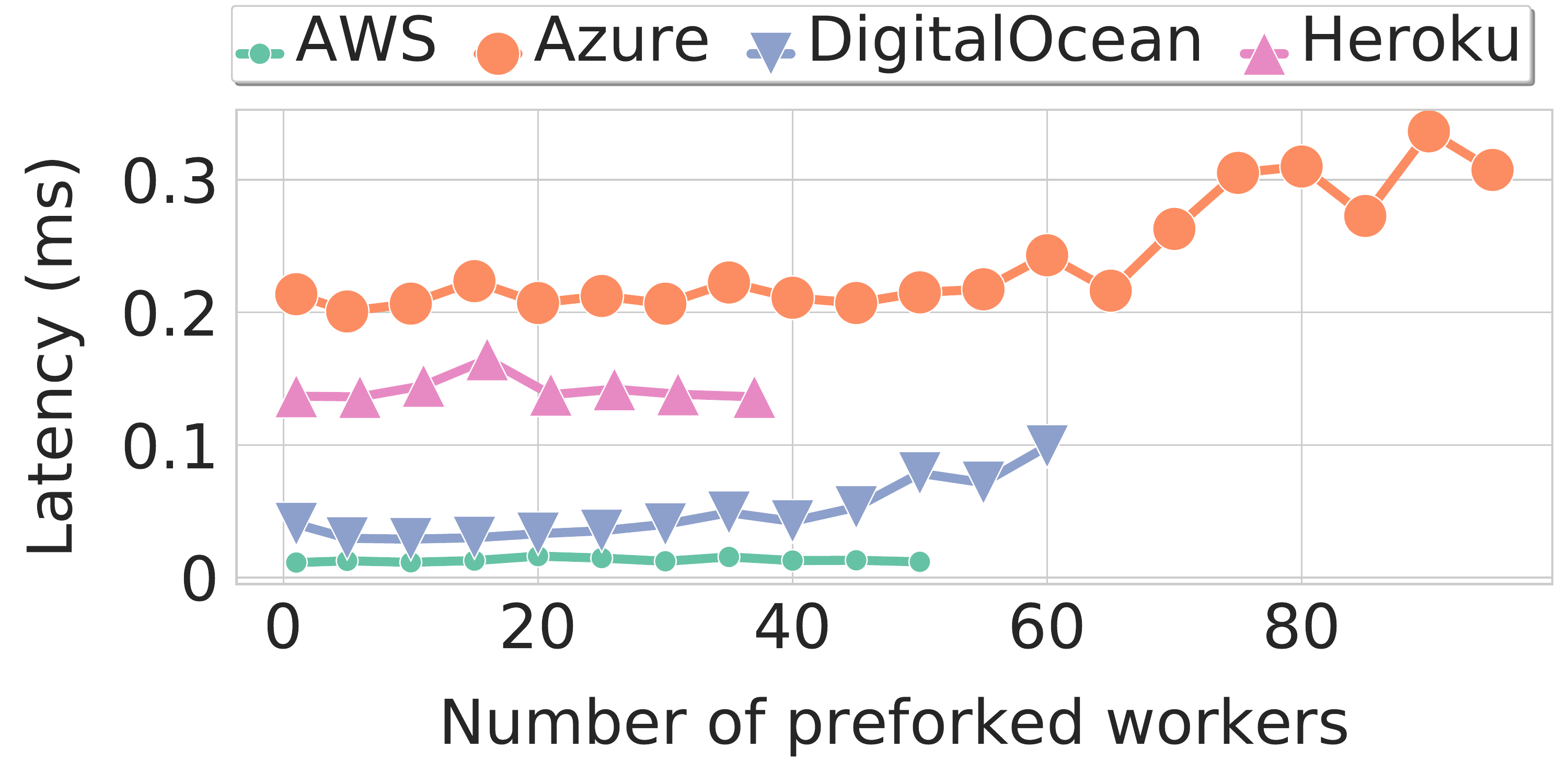}
\label{fig:workers_vs_latency}}
\hspace{0.1cm}
\subfigure[Attacker's gain comparison for 50ms payloads and different attack's bandwidths, against the best performing Node.js setup from Section~\ref{res:mitigation}, for each PaaS platforms.]{
\includegraphics[keepaspectratio=true,angle=0,width=.3\linewidth] {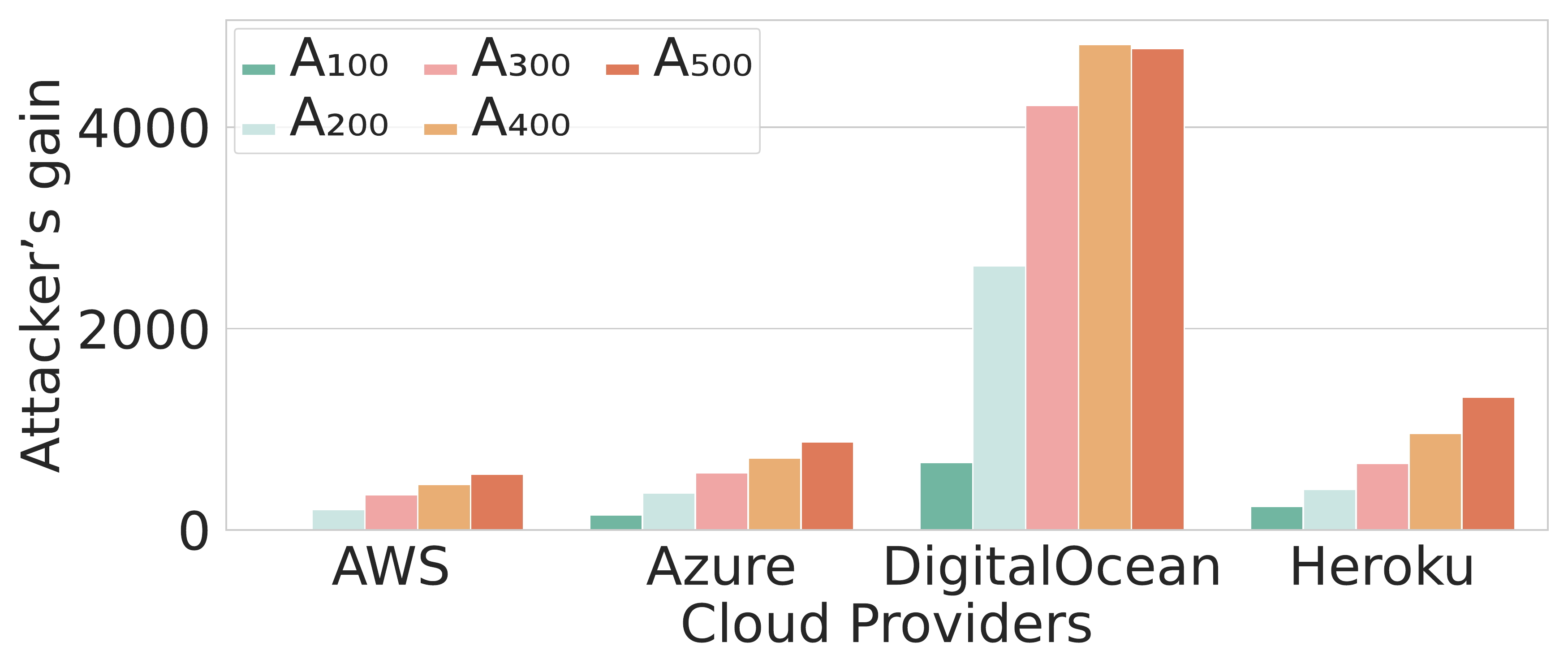}
\label{fig:different_attack_models}}
\vspace{-2mm}
\caption{Impact of increasing the number of preforked workers on a) Attacker's gain and b) latency in non-attack conditions in Node.js. Figure c) shows the comparison of Attacker's gain for \textit{50ms} payloads in different attack's bandwidths on each PaaS platforms.}
\label{fig:workers_vs_gain_and_latency}
\vspace{-5mm}
\end{center}
\end{figure*}

For event-based systems, we propose a much cheaper mitigation strategy: we hypothesize that by increasing the number of preforked workers, one can achieve a better resilience to CPU-based DoS attacks at a low cost. The current developers' consensus is that the number of workers should be equal to (twice) the number of physical cores available on the machine\footnote{\url{https://docs.gunicorn.org/en/stable/design.html\#how-many-workers}}\footnote{\url{https://devcenter.heroku.com/articles/node-concurrency\#tuning-the-concurrency-level}}. In our local experiments described in Appendix~\ref{apx:workers}, we find that increasing this value beyond this recommendation provides increased resilience to the attack. To see if this hypothesis holds in the cloud, for each platform, we rerun our Node.js experiments with increasing number of threads. Since most platforms limit the number of workers to small values, we consider five configurations between 1 and 20, using increments of five. In
Figure~\ref{fig:workers_vs_gain}, we show the \emph{Attacker's gain} for all configurations and all cloud providers, for Node.js versus the default Tomcat configurations. When comparing with Figures~\ref{fig:aws_experiment}-\ref{fig:digitalocean_experiment}, we see that the default configurations are clustered at the top of the figure and that most configurations with a higher number of workers have lower \emph{Attacker's gain}, comparable to some of the Tomcat servers. In AWS, the gain for Node.js with 20 workers drops by 2,222 from the default settings. However, as we discuss in Appendix~\ref{apx:examples}, the responses of a Node.js  and a Tomcat server with comparable \emph{Attacker's gain}s, are very different.

While this approach is effective at increasing the server's resilience to the attack, it also comes with a cost. Figure \ref{fig:workers_vs_latency} shows the latency of benign requests, in non-attack conditions, for different numbers of workers, on the four cloud platforms. Increasing the number of workers has a negative effect on the latency of requests. While the latency increases by 30\% for the maximum number of workers in Azure, it increases by 59\% in DigitalOcean. Thus, even though fixing up the number of workers to a high value improves the resiliency of the systems against DoS attacks, the system will suffer from higher latency in non-attack scenarios. %Secondly, it is not trivial find the optimal number of workers for  a Node.js system which provide resonable resiliency against DoS attacks while keeping the benign request latency minimum. There are multiple local minima in the latency graph for Heroku and AWS which indicates seeting up the optimal number of workers is a optimization problem. This calles for a dynamic solution for setting up the optimal number workers depending on the attack scenario. Developing such a dynamic solution for Node.js and other architectural systems is out of the scope of this work and we leave it as our future work.

\iffalse

\begin{figure}
    \centering
    \includegraphics[keepaspectratio=true,width=0.7\linewidth]{figures/different_attack_model.pdf}
    \caption{Attacker's gain comparison for 50ms payloads and different attack's bandwidths, against the best performing Node.js setup from Section~\ref{res:mitigation}, for each PaaS platforms.}
    \vspace{-5mm}
    \label{fig:different_attack_models}
\end{figure}

\fi

\subsection{Q3: Vulnerability threshold}
\label{res:threshold}
Our results so far show that there is no special threshold from which a CPU-heavy payload should be considered a vulnerability. Instead, we observe a clear correlation between the size of the payload and the attacker's advantage during the attack. For example, using 500 ms payloads, which are not considered problematic by practitioners, in Node.js on Heroku, the attackers can bring down the server for more than 50s, which is more than 90\% of the considered measurement window. We now explore a previously neglected part of the attack model: can the attacker use larger bandwidths with small payloads for an effective attack? To this end, for each platform, we choose the best performing setup in our previous experiment, i.e., the number of preforked workers that gives the lowest total gain, and we simulate attacks with small payloads against it.

Figure \ref{fig:different_attack_models} shows the result of attacking these servers with 50ms payloads, using increasing attack's bandwidth. 
In all four platforms, it is possible to achieve a higher gain using a larger bandwidth, which is not entirely unexpected. However, it is once again surprising that there is no built-in mechanism to prevent this increase in bandwidth, e.g., rate-limiting. Another unexpected effect is that by increasing the bandwidth five times, we obtain an average 16-fold increase in \emph{Attacker's gain}. %comparable to the gain obtained for 100ms payloads.
While the server never goes completely down in the smallest bandwidth attack, the throughput falls below 10\% for five, seven, 48, and 12 seconds in AWS, Azure, DigitalOcean, and Heroku respectively, in the case of the largest bandwidth attack. We conclude that very small payloads can also be leveraged to harm web applications.
\vspace{-2mm}

\subsection{Discussion}
\label{sec:discussion}
% Load balancer between the thread pool ad the workers pool
Considering the trade-off discussed in Section~\ref{res:mitigation} between the server's resilience to CPU-based DoS attacks and its latency in non-attack conditions, we propose a load balancing solution for event-based systems: dynamically adjust the sizes of the asynchronous I/O thread pool and of the (preforked) workers pool, based on the type of received requests. That is, if a lot of I/O-heavy requests are observed, the thread pool should be reasonably high, but if the requests are CPU-heavy, the number of workers should be high instead. We believe that by implementing this strategy in the current event-based systems, one can increase their resilience to CPU-based DoS, without the additional cost for the non-attack scenario.

% Reuse methodology for penetration testing
The results in Section~\ref{res:threshold} show that small payloads can be weaponized against well-configured servers in the cloud. Hence, we propose a multi-step mitigation strategy. First, developers should strive to limit the attacker's bandwidth as much as possible, e.g., by using aggressive rate limiting. Second, they should identify any slow computation in their web server, i.e., as low as few tens of milliseconds. As discussed earlier, something as widely used as decryption with a strong key may be of concern. Developers should be encouraged to make slow computation hard to trigger, e.g., only allow it to logged-in users. Using the largest identified CPU-heavy request and the maximum bandwidth, developers should estimate the maximum damage an attacker can achieve against their server, possibly using our methodology.

%A different aspect worth mentioning is that our methodology can be used by security analysts to estimate the impact of CPU-based DoS attack against a target system. To this end, they must first identify API endpoints that perform suspiciously slow. Then, they should simulate an attack by sending multiple requests invoking that particular endpoint, and simultaneously, perform measurement requests. By analyzing the collected data using the  metrics we proposed, the analyst can evaluate the feasibility and the  impact of the attack, and thus, take informed decisions about the required security controls.

% Use the new metrics to optimize a defense/attack
%Moreover, by repeating the above process multiple times, developers can also solve several optimization problems, using the values produced by our metric. For example, they can search the attack parameters that maximize the attacker's gain, or identify the best configuration for the intrusion detection system that minimizes the throttled time. As discussed earlier, we do not recommend optimizing for a single metric, but to perform multi-objective optimization, instead.

% Move the threshold discussion here?

% Limitations

\section{Related work}

\paragraph{CPU-based denial-of-service attacks}
Crosby and Wallach~\cite{crosby_denial_2003} were the first to introduce bandwidth-bound, CPU-based DoS attacks. They analyze how an attacker can utilize vulnerabilities in hash tables and binary trees to deteriorate the performance of different systems. Shan et al.~\cite{shan_tail_2017} show another variety of low-volume DoS attacks that utilize the architectural vulnerabilities in n-tier web applications, with repetitive small attack windows. Meng et al.~\cite{meng_rampart_2018} propose Rampart, a defense against CPU-exhaustion attacks that uses function profiling to detect and stop attacks at run time. Demoulin et al.~\cite{demoulin_detecting_2019} propose a machine learning model that uses the resource monitoring system in Linux kernel to detect long-running requests at run time. REGEXNET~\cite{bai_runtime_2021} uses a machine learning-based system to classify requests as malicious, and run them in a sandbox as a defense. Davis et al.~\cite{DavisWL18} propose timeouts as a way to prevent triggering CPU-heavy payloads.
None of this work analyzes the effect of the server's architecture on its resilience to DoS attacks. 

\paragraph{Algorithmic complexity vulnerabilities}
%Low volume algorithmic complexity vulnerabilities attacks received significant attention from both attackers and researchers because of the stealthiness and effectiveness of the attack. 
There is an extensive body of work to detect and mitigate algorithmic complexity vulnerabilities, most of them focusing on a particular system or application. Various dynamic~\cite{liu_revealer_2021} and static program analysis~\cite{chang_inputs_2009, static_wustholz_2017} tools have been proposed for detecting these security problems. Olivo et al.~\cite{olivo_detecting_2015} introduce second-order DoS vulnerabilities and propose a static analysis technique for  detecting them. Previous work also discusses how bugs in serialization libraries~\cite{DietrichJRTP17} or parsers~\cite{Rasheed0T21} can be used to exhaust CPU resources of a system.
Shen et al.~\cite{shen_rescue_2018}, Davis et al.~\cite{case_davis_2017, davis_using_2021}, and Staicu et al.~\cite{staicu_freezing_2018} study regular expressions vulnerabilities, a widespread problem in the Node.js ecosystem. DISCOVER~\cite{awadhutkar_discover_2019} analyzes Java source code and detect vulnerabilities. REVEALER~\cite{liu_revealer_2021} uses both static and dynamic analysis to detect vulnerable regular expressions, and then produces an input string to exploit the detected vulnerability. Researchers have also proposed different fuzzing-based~\cite{petsios_slowfuzz_2017, blair_hotfuzz_2020} or symbolic execution-based~\cite{NollerKP18,BurnimJS09} systems to automatically generate malicious inputs that trigger algorithmic complexity vulnerabilities. Vulnerabilities can also be present in third-party code, published in open-source ecosystems~\cite{DavisCSL18,ZimmermannSTP19,AlfadelCS21}. Instead of studying algorithmic complexity vulnerabilities directly, we study their potential impact on web applications deployed in the cloud.

% \vspace{-4mm}
\paragraph{Performance of web servers}
%Although there are several comparative studies on performance of different web server architectures, they mostly focus on typical usage scenarios for a web application, while we are interested in the case when an adversary aims to repeatedly send easy-to-trigger CPU-heavy computations.
Pariag et al.~\cite{pariag2007comparing} compares the performance of event-driven, thread per connection, and staged event-driven architectural servers. They also propose a new high-throughput architecture incorporating their suggested modification. Pai et al.~\cite{pai1999flash} propose an asymmetric, multi-process, event-driven architecture for web servers that combines the performance of both event-driven and multi-threaded servers. Von Behren et al.~\cite{von2003events} show a performance comparison of event-driven and multi-threaded architectures and demonstrated that by modifying the default threading implementation, multi-threaded servers can also handle large concurrent requests. While related, none of this work deals with malicious adversaries.
Millibottlenecks~\cite{PuKLZLPWJXMWJKS17,WangLKZP17} are a known performance problem of web applications that can lead to long-tail latency problems. Shan et al.~\cite{shan_tail_2017} show how an adversary may leverage this to attack web applications, and Zhange et al.~\cite{ZhangSWLYW19} show that collocating virtual machines in the cloud can also produce long-tail effects in web applications. We are the first to study the effect of equivalent DoS attacks on the performance of widely-used web servers.

%\paragraph{Long-tail latency}
%\crs{\cite{ZhangWKSH20,ZhangSWLYW19,TennagePJJ19,BergerBZ0H18,KarstenB20,YangSLYC18}}

\paragraph{Performance problems}
%Finding the performance issues in a program is very critical for mitigating algorithmic complexity vulnerabilities attack. 
A lot of work have been done for detecting performance bottlenecks, profiling programs, and generating problematic input. 
Wei et al.~\cite{wei_singularity_2018} propose a genetic algorithm-based program synthesis that works on recurrent computation graphs to find the worst-case asymptotic complexity of an application. PerfFuzz~\cite{lemieux_perffuzz_2018}  automatically generates inputs that trigger the worst-case performance of a program, using a multi-directional feedback loop. PerfSyn~\cite{toffola_synthesizing_2018} uses combinatorial and graph search algorithms to synthesize programs to find performance bottlenecks, while SyncProf \cite{yu_syncprof_2016} uses concurrency-focused profiling to detect synchronization bottlenecks in programs. Nistor et al.~\cite{NistorSML13} propose an oracle based on  memory access patterns for detecting performance bugs.
Selakovic et al.~\cite{selakovic_performance_2017} identify the root causes of performance issues in popular JavaScript projects, and He et al.~\cite{HeJLXYYWL20} propose a testing framework for identifying configuration-related performance problems. Liu et al.~\cite{LiuXC14} study performance bugs in Android applications and propose a static analysis approach for detecting them. While performance problems can be used for DoS attacks, none of the work above considers malicious adversaries. \mas{ Significant work has also been done on server scalability, performance, and availability~\cite{wolter2010performance, manjhi2006simultaneous, alomari2012autonomic, nylander2018cloud, tadakamalla2020autonomic}, but none of them focuses on malicious adversaries that aim to repeatedly trigger slow computations.}
%avritzer2010monitoring
%bhatti1999web

\paragraph{Comparison of ecosystems and frameworks} 
Duan et al.~\cite{duan_towards_2021} use program analysis techniques to assess the functional and security features of package managers in Python, Node.js, and Ruby.
Kikas et al.~\cite{10.1109/MSR.2017.55} also report the significant differences in package dependency networks across three ecosystems.
%Vasilakis et al.~\cite{vasilakis2021supply} propose HARP to generate input and infer the output of string-processing components in JavaScript and C/C++ using active library learning and regeneration techniques. 
Decan et al.~\cite{decan2017empirical} analyzes the package dependency networks for seven languages and proposes novel metrics to highlight the growth, similarity, and differences between networks.  We are the first to compare the resilience of widely-used web servers to CPU-based DoS attacks.

%\paragraph{Vulnerabilities in scripting languages}
\section{Conclusion}
In this work, we study the resilience of widely-used web servers to CPU-based denial-of-service attacks. We show that servers react very differently to attacks: event-based systems tend to exhibit a very sudden drop in quality of service, while one-thread-per-client systems suffer more gradual performance degradation. We also show that, contrary to practitioners' beliefs, sub-second slowdowns can be leveraged for effective attacks against web applications deployed in the cloud, under realistic deployment conditions. Our results are a call to arms for new defenses against CPU-based DoS attacks, and a warning about the hidden security trade-off involved in choosing the architecture of a web application.

\bibliographystyle{ACM-Reference-Format}
\bibliography{references}
\appendix

\section{Ethics}
This work does not raise any ethical concerns, as all our experiments were carried against servers under our control.

\section{Multiple workers in local setup}
\label{apx:workers}
\begin{figure}
    \centering
    \includegraphics[keepaspectratio=true,width=.7\linewidth]{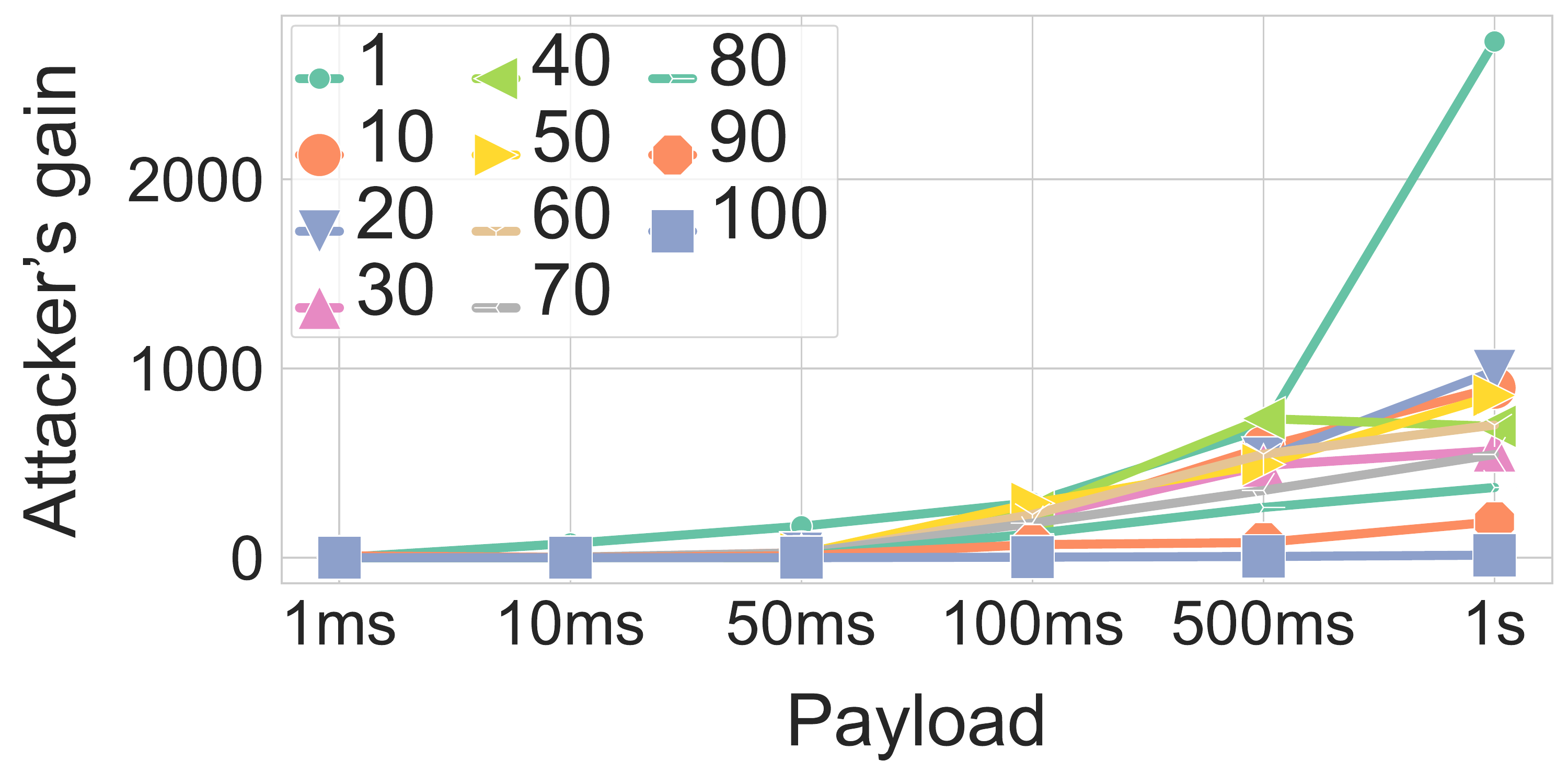}
    \caption{Attacker's gain for Node.js with different number of workers in local setup.}
    \label{fig:local_gain}
\end{figure}

To test the hypothesis that high number of preforked workers can increase the resilience of event-based systems to CPU-based DoS attacks, we perform a local experiment with Node.js. Using the built-in \texttt{Cluster API}, we vary the number of workers in increments of ten, and simulate attacks with different payloads, against each configuration. In Figure~\ref{fig:local_gain}, we show the \emph{Attacker's gain} for various numbers of workers. The results show that there is an inverse relationship between the number of workers and the \emph{Attacker's gain}: the higher the number of workers, the more resilient the server is to an attack.

\begin{figure*}
\begin{center}

\subfigure[Apache in Azure]{%\protect\footnotemark
\includegraphics[keepaspectratio=true,angle=0,width=.45\linewidth] {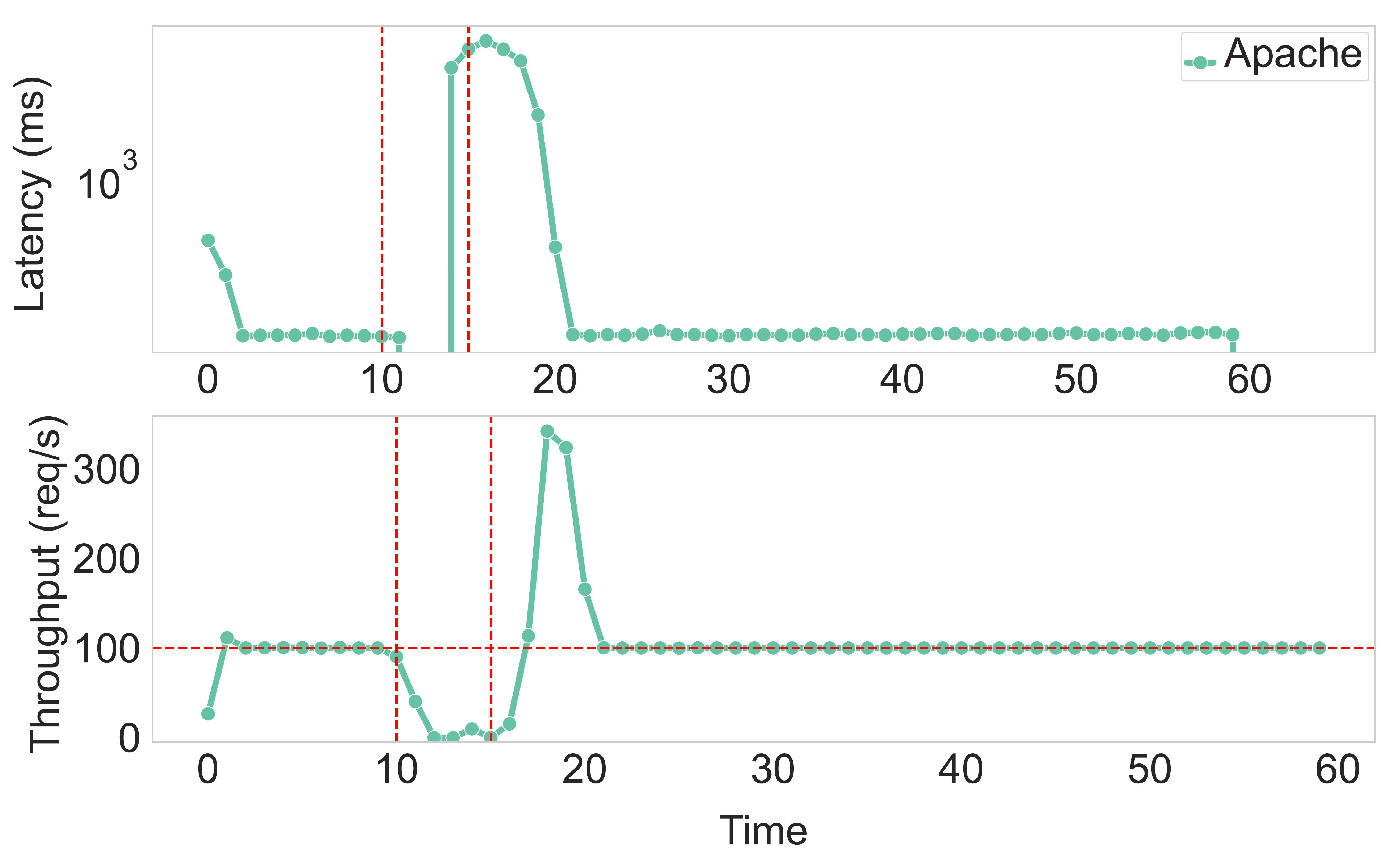}
}
\hspace{0.1mm}
\subfigure[Apache in Heroku]{
\includegraphics[keepaspectratio=true,angle=0,width=.45\linewidth] {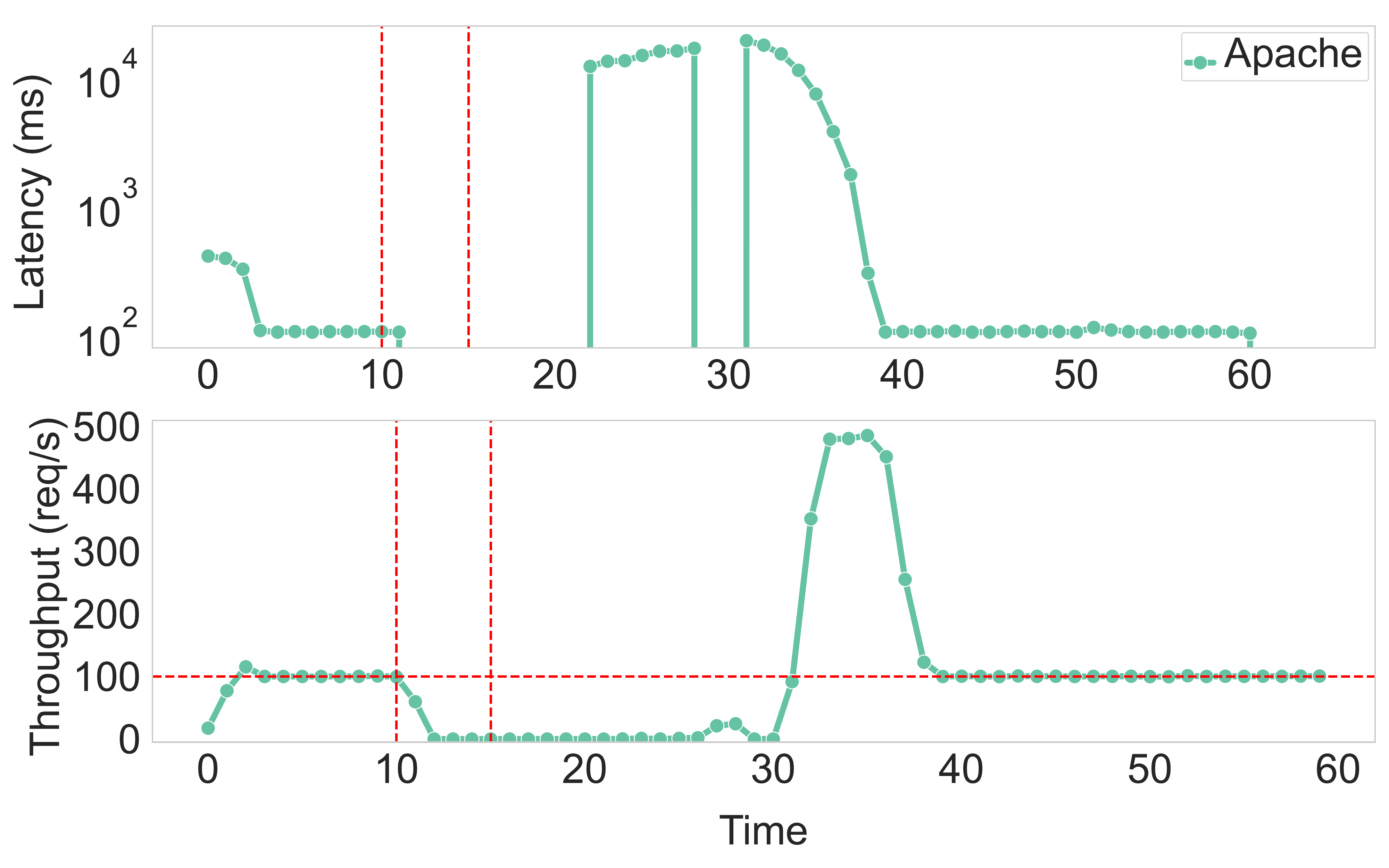}}
%\vspace{-2mm}
\caption{Response of two different Apache servers to the same attack, i.e., $\mathcal{A}=(100, 500ms, 5s)$.}
%\vspace{-2mm}
\label{fig:php_two_setups}
\subfigure[Node.js]{%\protect\footnotemark
\includegraphics[keepaspectratio=true,angle=0,width=.45\linewidth] {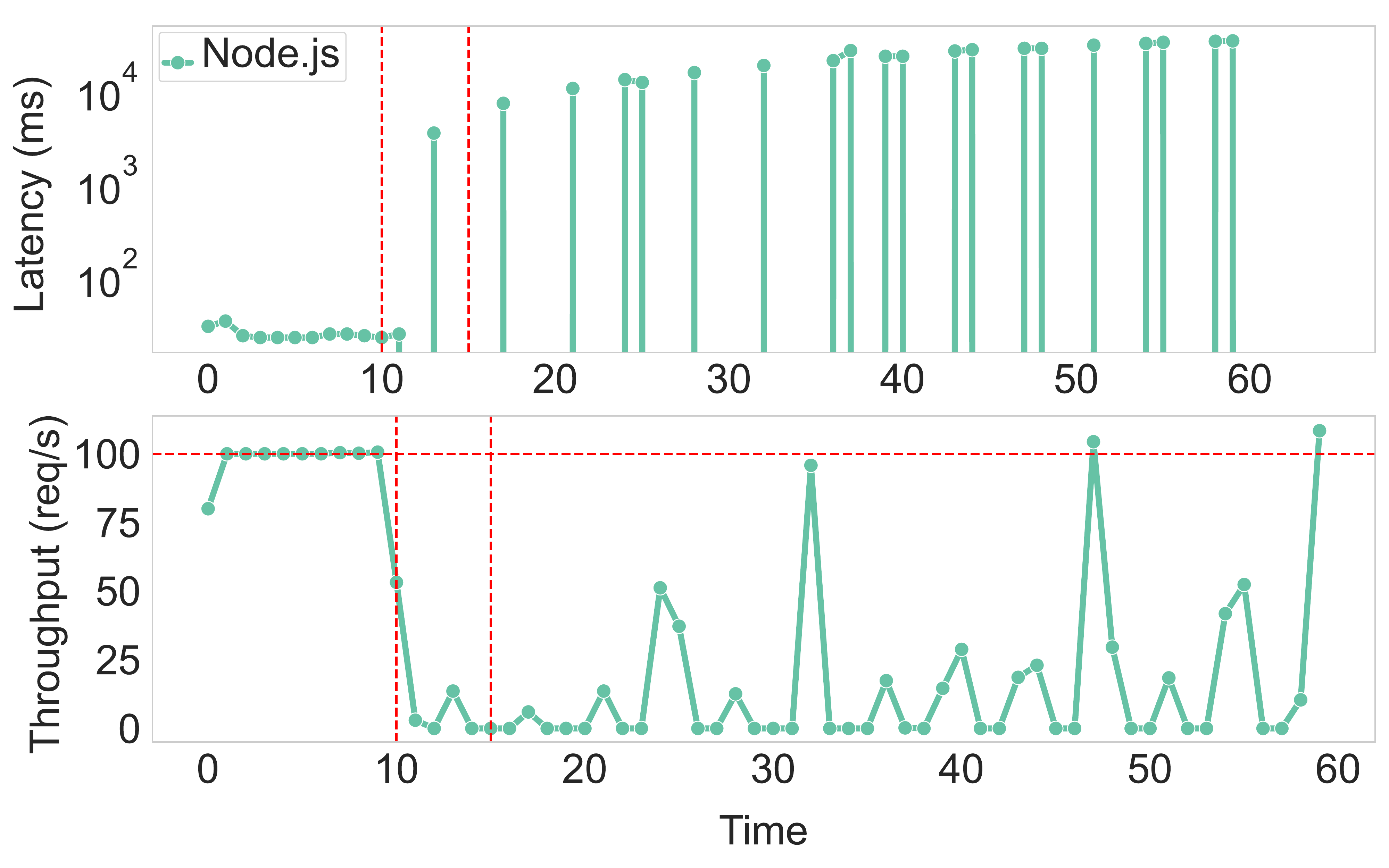}
}
%\hspace{0.2mm}
\subfigure[Tomcat]{
\includegraphics[keepaspectratio=true,angle=0,width=.45\linewidth] {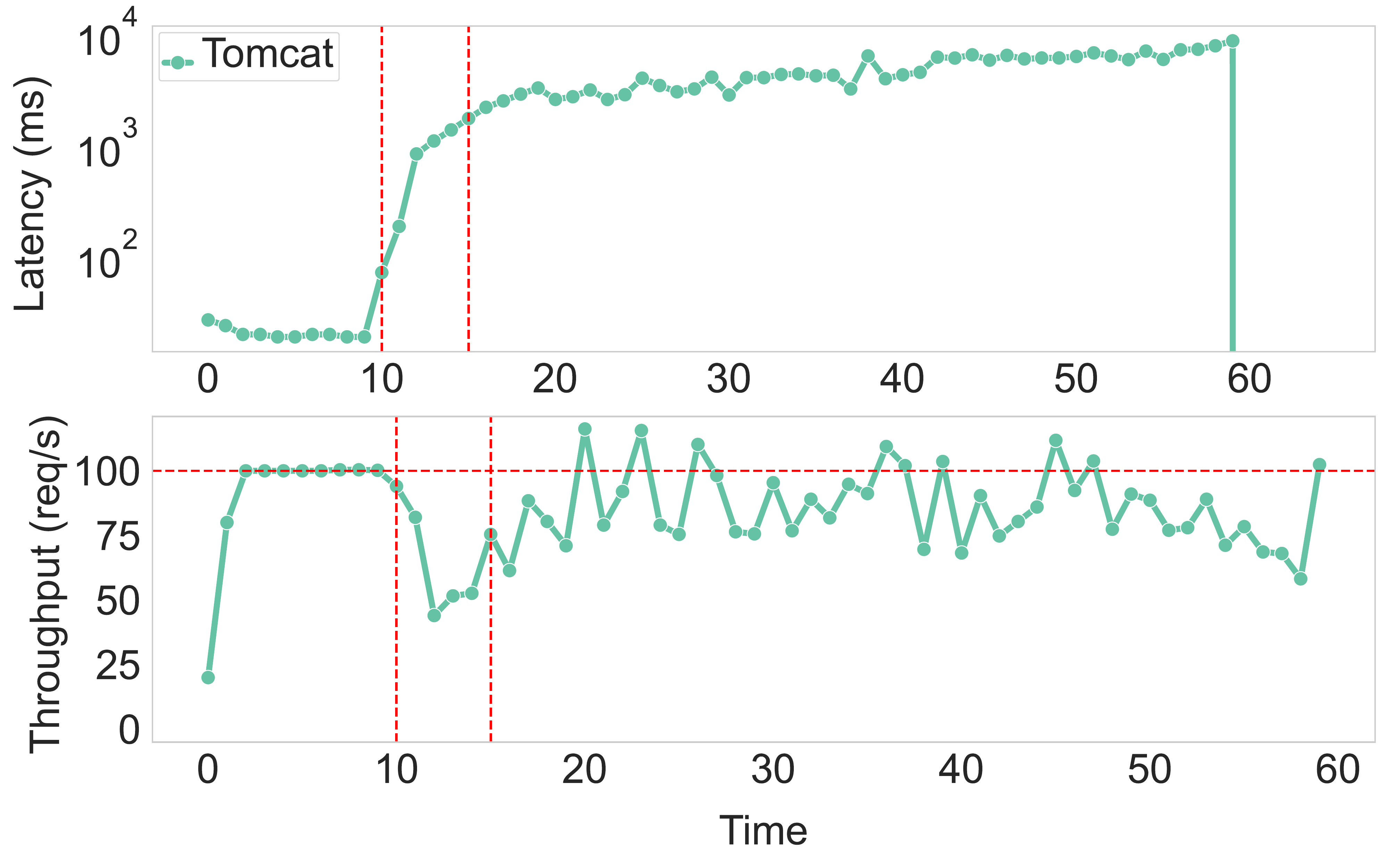}
}
\subfigure[Gunicorn]{
\includegraphics[keepaspectratio=true,angle=0,width=.45\linewidth] {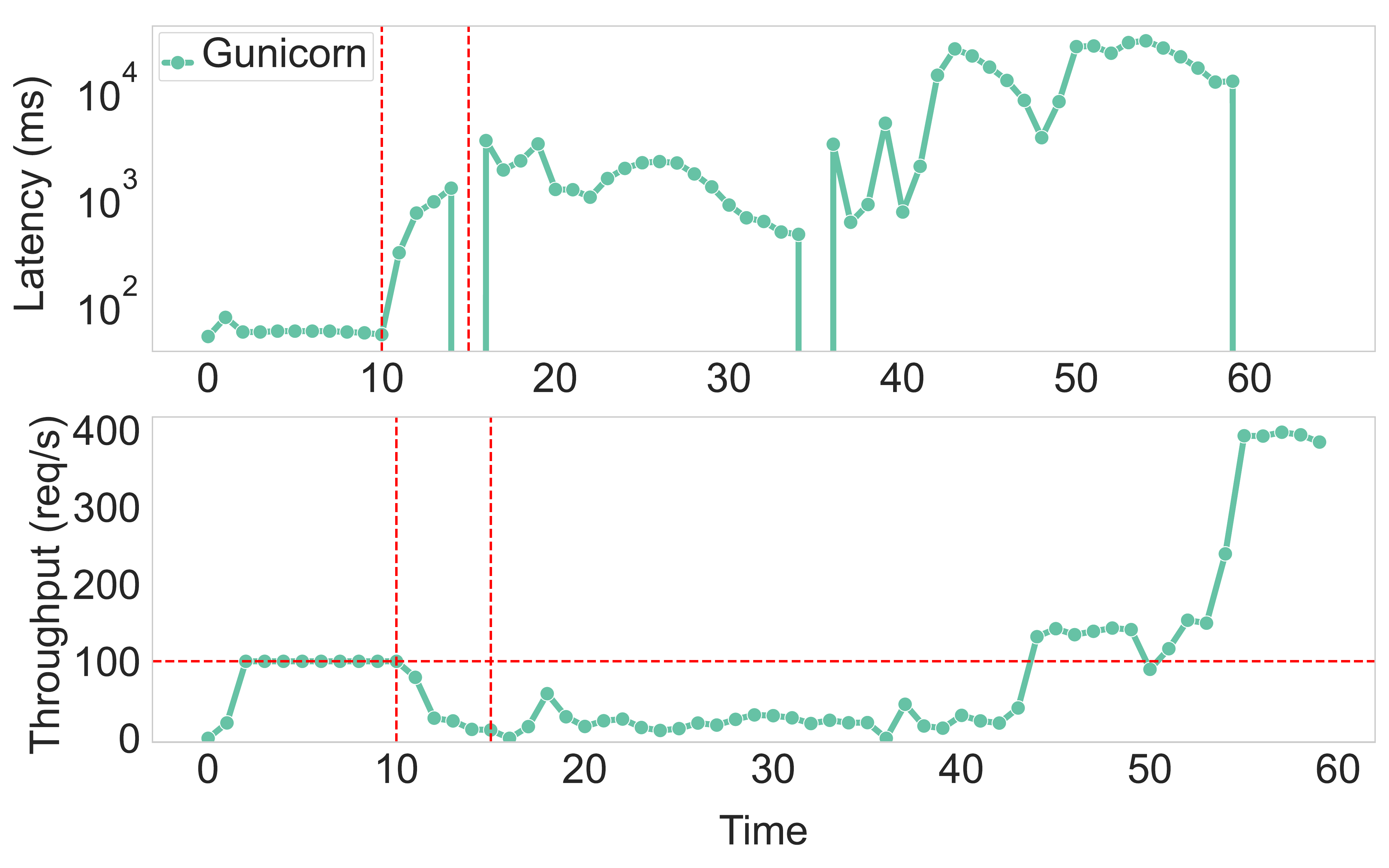}
}
\subfigure[Go]{
\includegraphics[keepaspectratio=true,angle=0,width=.45\linewidth] {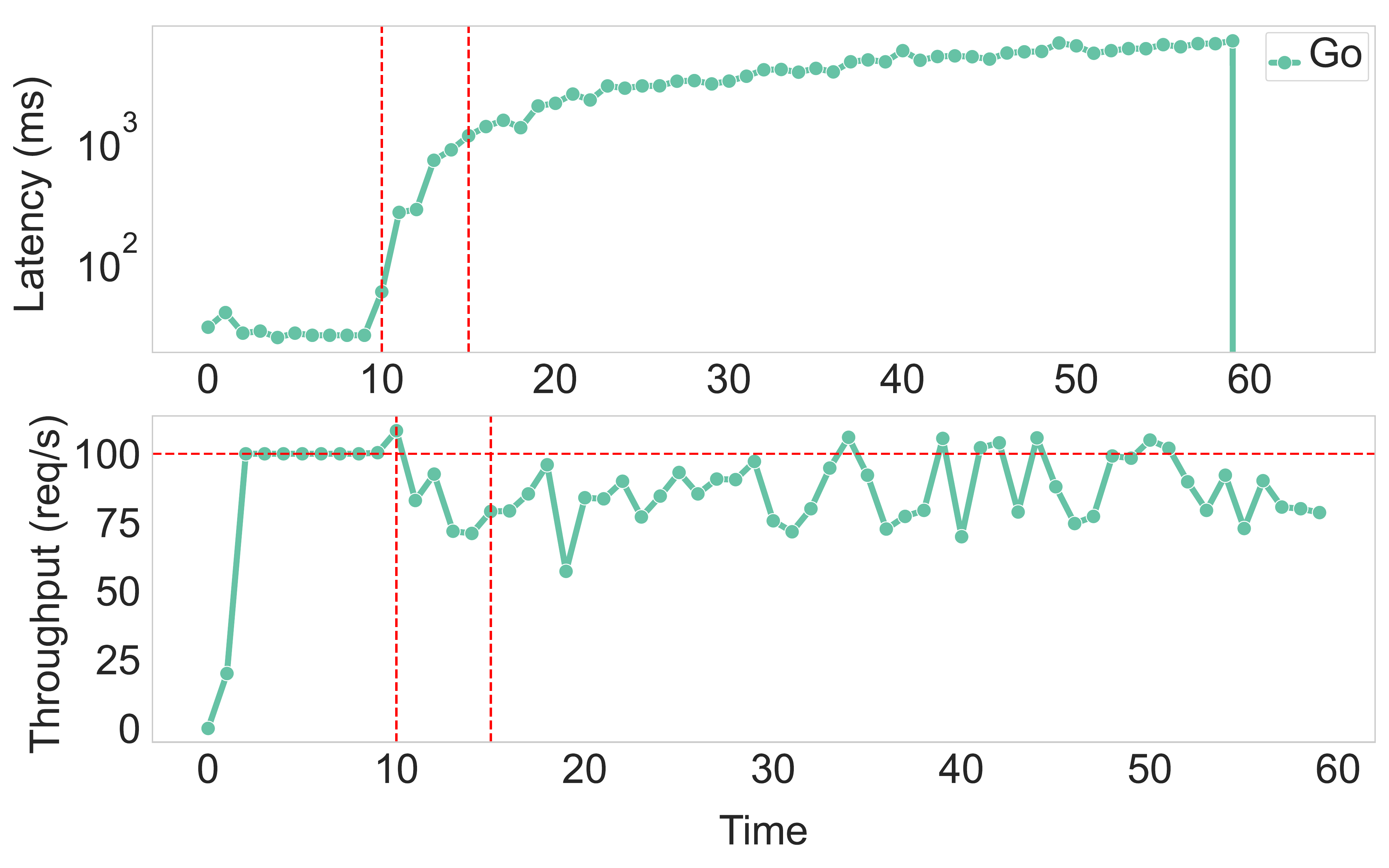}
}
%\vspace{-2mm}
\caption{Response of different servers to the same attack, i.e., $\mathcal{A}=(100, 1s, 5s)$, in DigitalOcean.}
\label{fig:do_multiple}
\end{center}
\end{figure*}

\begin{figure*}
\begin{center}

\subfigure[Tomcat in Azure]{%\protect\footnotemark
\includegraphics[keepaspectratio=true,angle=0,width=.45\linewidth] {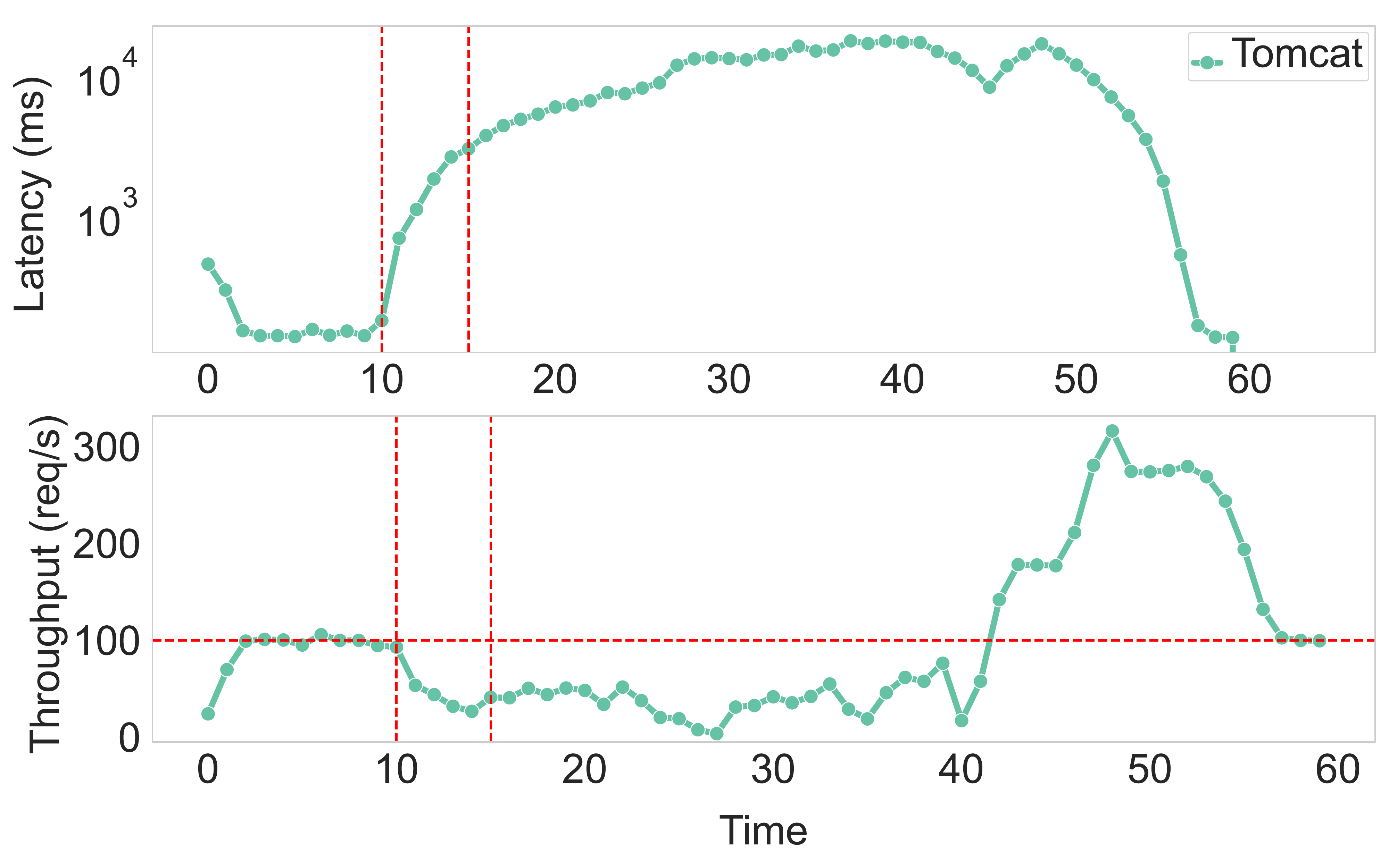}
}
\hspace{0.1mm}
\subfigure[Node.js in AWS, with 20 preforked workers]{
\includegraphics[keepaspectratio=true,angle=0,width=.45\linewidth] {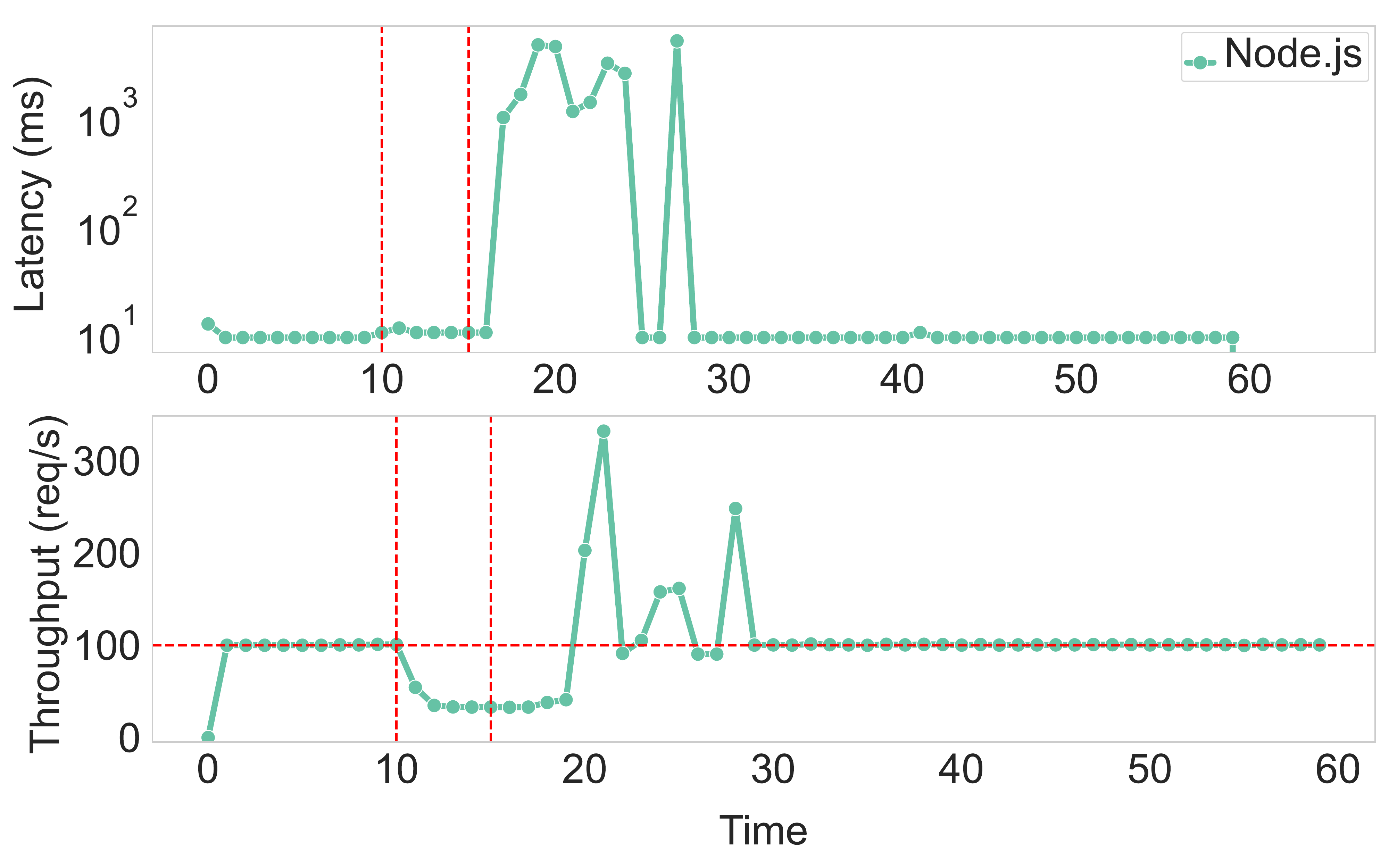}
}
%\vspace{-2mm}
\caption{Response of two different servers to the same attack, i.e., $\mathcal{A}=(100, 1s, 5s)$.}
\label{fig:node_vs_spring}

\end{center}
\end{figure*}
\section{Example responses to attack}
\label{apx:examples}
%While overall, the system on the right exhibits lower attacker's gain than the one on the left, e.g., the drop in throughput is smaller, the one on the left has a worst short-term degradation between the 10th and the 20th second. On the contrary, the system on the right recovers much fsater
Below, we discuss several interesting server responses that we encountered during our experiments. 

Let us first consider Figure~\ref{fig:php_two_setups}, showing Apache's response to 500 milliseconds payloads in two different cloud providers. For both setups, there are several seconds with zero throughput, in which no requests are served. There is also a high difference in recovery time between the two setups: a couple of seconds after the attack stopped in Azure, and more than ten seconds in Heroku. This could, in theory, be caused by the better hardware used in Azure instances, but the difference is too high to be explained in this way.  It is also interesting to compare these results with Apache's response to the same attack in our local setup, showed in Figure~\ref{fig:local_overview}. Seeing all these dramatic differences, we hypothesize that Apache runs using different multi-processing modules\footnote{\url{https://httpd.apache.org/docs/2.4/mpm.html}}, in all these setups. This could explain the inconsistent responses we see for this web server.

%\footnotetext{Contrary to what many readers may think, the top figure does not depict a boa constrictor digesting an elephant, but the actual latency of the server during the attack.}

In Figure~\ref{fig:do_multiple}, we see the response of four servers to an attack with five seconds payloads, in DigitalOcean. Gunicorn appears to recover the fastest, i.e., the throughput consistently exceeds the expected one 25 seconds after the attack. For Tomcat and Go, while the throughput is kept at consistent levels throughout the experiments, the latency is continuously increasing during our observation period. Hence, we argue that it is hard to decide which one of these behaviors is the desired one for a web server. Another interesting aspect worth noticing is the high number of spikes we observe for Node.js, Tomcat, and Go. We do not observe this effect in any other setup we consider. In most of the other cases, we obtain a rather smooth throughput curve, like the ones in Figure~\ref{fig:php_two_setups} or Figure~\ref{fig:node_vs_spring}. We hypothesize that the cause for this effect in DigitalOcean is server-independent, e.g., the load balancer scheduling requests in a peculiar order. 

Finally, let us discuss the examples Figure~\ref{fig:node_vs_spring}. While Tomcat and Node.js are deployed on two different cloud providers here, i.e., Azure and AWS, respectively, we argue that the underlying hardware is similar as seen from Table\ref{architecure_info}. For the considered attack and cloud setups, we can safely say that Node.js performs better than Tomcat: the drop in throughput is smaller for Node.js, and it lasts for a shorter time, while the increase in latency is only perceivable for 20 seconds, as opposed to 40 seconds for Tomcat. Nevertheless, we observe that the increase in latency is gradual for Tomcat but sudden for Node.js. This example illustrates that event-based servers do not always perform worse than thread-based ones against CPU-based DoS. However, it is unlikely that developers use such high numbers of preforked workers in practice.
\iffalse
\appendix

\section{Stability of the measurements}
\begin{table}
\center
\scriptsize
\setlength\tabcolsep{3pt}
\caption{
  Stability of repetead throughput measurements performed under the same attack conditions. The stability is calculated using root-mean-square deviation.
}
\begin{tabular}{lllll}
\toprule
Setup & Web server & \multicolumn{3}{c}{Pairwise stability} \\
& & Minimum & Median & Maximum \\
\midrule
Local setup & Spring & \\
 & PHP & \\
 & Django & \\
 & Go & \\ 
 & Node.js & \\ 
\midrule 
AWS & Spring & \\
 & PHP & \\
 & Django & \\
 & Go & \\ 
 & Node.js & \\
\midrule 
Azure & Spring & \\
 & PHP & \\
 & Django & \\
 & Go & \\ 
 & Node.js & \\ 
\midrule 
Digitalocean & Spring & \\
 & PHP & \\
 & Django & \\
 & Go & \\ 
 & Node.js & \\  
\midrule 
Heroku & Spring & \\
 & PHP & \\
 & Django & \\
 & Go & \\ 
 & Node.js & \\  
\bottomrule
\end{tabular}
\end{table}
\fi
\end{document}